\documentclass[a4paper]{jpconf}
\usepackage{graphicx}
\usepackage{psfrag} 

\newcommand{\be}[0]{\begin{equation}}
\newcommand{\ee}[0]{\end{equation}}
\newcommand{\bea}[0]{\begin{eqnarray*}}
\newcommand{\eea}[0]{\end{eqnarray*}}

\begin{document}
\begin{center}
{\large{\bf Evolving images of the proton:\\[3mm] Hadron physics over the past 40 years}}
\end{center}
\vspace{3mm}

\begin{center}
{\bf Michael R. Pennington}

Thomas Jefferson National Accelerator Facility, 

12000 Jefferson Avenue, Newport News, VA 23606, USA
\end{center}
\vspace{3mm}
\begin{center}
 {\bf Abstract}
\end{center}

\vspace {3mm}
Once upon a time, the world was simple:  the proton contained three quarks, two {\it ups} and a {\it down}. How these give the proton its mass and its spin seemed obvious. Over the past forty years the proton has become more complicated, and how even these most obvious of its properties is explained in a universe of quarks, antiquarks and gluons remains a challenge. That this should be so should come as no surprise. Quantum Chromodynamics, the theory of the strong interaction, is seemingly  simple, and its consequences are straightforward in the domain of hard scattering where perturbation theory applies. However, the beauty of the hadron world is its diversity. The existence of hadrons, their properties, and their binding into nuclei do not appear in the Lagrangian of QCD. They all  emerge as a result of its strong coupling. Strong coupling QCD creates complex phenomena, much richer than known 40 years ago: a richness that ensures colour confinement and accounts for more than 95\% of the mass of the visible Universe. How strong coupling QCD really works requires a synergy between experiment and theory. A very personal view of these fascinating  developments in cold QCD is presented.

\newpage
\section{In the beginning}
\parskip=2mm
The strong force goes back to the beginning of time. Our understanding of how this has shaped the Universe, from creating nuclear matter to building stars is, of course, much more recent. Nevertheless, our understanding has evolved dramatically over the 100 years since the atomic nucleus was first revealed. Twenty years later with the discovery of the neutron, it soon became clear that the nucleus contained both charged and neutral constituents, protons and neutrons collectively called nucleons. Yukawa proposed~\cite{yukawa} that these were bound by the exchange of a much lighter particle, the pion, later found in cosmic rays~\cite{powell}, the mass of which determines the range of nuclear forces.  Then followed further discoveries first from cosmic rays (for instance~\cite{rochester}) and then from the first accelerator experiments: protons, neutrons and pions were not alone inside the nucleus. Emerging was a whole universe (the femto-universe) inside each nucleus swirling with many strongly interacting particles, we call hadrons. Indeed, these many hadrons could be divided into two distinct categories: those with 1/2-integer spin, like the proton, forming the baryon family, and those with integer spin, like the pion, the meson clan.

However, fifty years ago the theory of the strong nuclear force was just a more complicated version of that of Yukawa: hadrons being exchanged generated the forces they experienced, not just the  pion, but all the mesons and baryons could be exchanged. No one particle was more fundamental than any other. There was nuclear democracy. Hadrons being exchanged in the crossed-channel generated hadronic resonances in the direct channel. They pulled each other up by their bootstraps.   Indeed, those with related quantum numbers, the same isospin and naturality, could be collectively exchanged as Regge trajectories~\cite{chew,frautschi}. Such trajectories related hadrons of different spin but the same flavor quantum numbers. In contrast, hadrons with the same spin and parity, but different flavours, formed quark model multiplets. Every known meson could be understood in terms of a quark and an antiquark, and baryons of three quarks~\cite{gell-mann,zweig}. 50 years ago the number of flavours was three: just  {\it up, down} and {\it strange}.
The {\it u} and {\it d} quarks seemed to have masses about a third of that of the proton, {\it viz.} $\sim 300$~MeV, and the {\it s} quark $\sim 150$~MeV bigger.  However these quarks were thought by many to be nothing more than mnemonics for the flavour properties of hadrons with no reality of their own. They were just a convenient accounting tool. In the past forty years, the pace of change of our world view of the nucleus and hadron  dynamics has quickened.

\begin{figure}[h]
\centering
\includegraphics[width=14.cm]{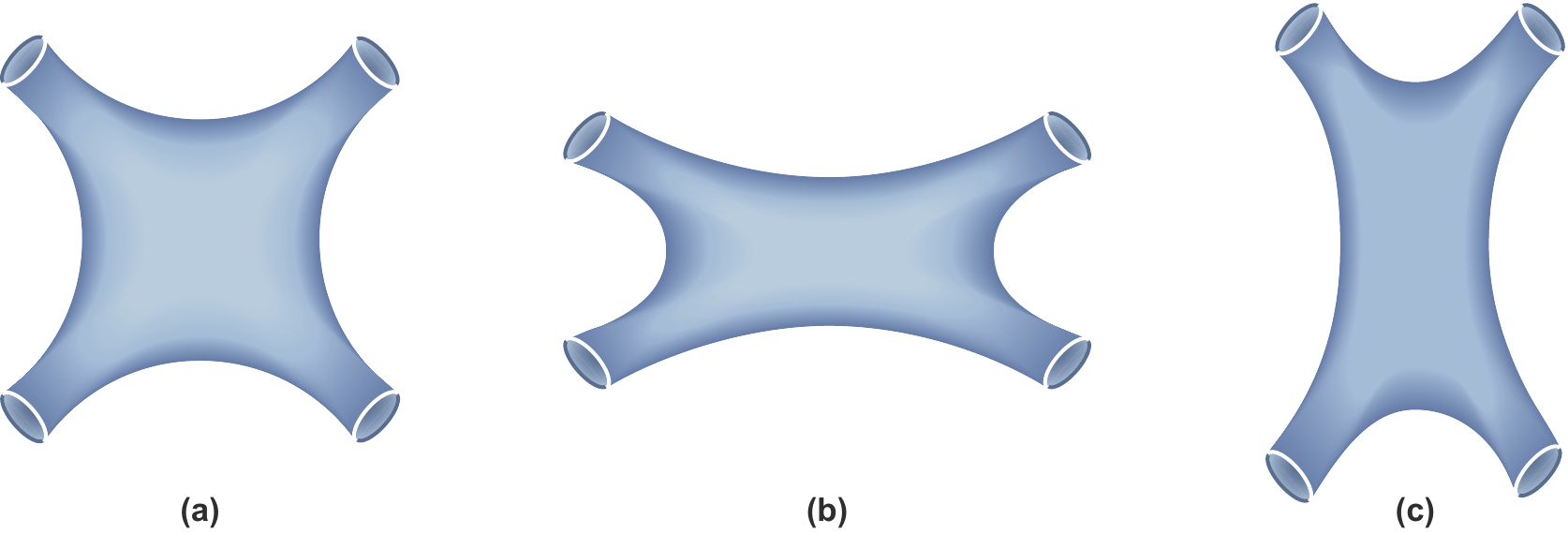}
\caption{$2\to 2$ scattering of mesons in a string picture of hadrons.  Mesons are oscillating strings with their flavour carried by quarks at the ends of the string. In the interaction the strings combne and then divide producing the world sheet shown. This picture combines direct channel dynamics, the appearance of resonances, with forces generated by meson exchanges in the crossed channel.
These can be seen to be encompassed by distorting the sheet horizontally or vertically. This picture is encoded in the explicit formula for the $2\to 2$ amplitude written down by Veneziano~\cite{veneziano}.  }
\label{fig-1}       
\end{figure}
Fifty years ago there was no microscopic theory of hadron dynamics. Other than protons, hadrons appeared as resonances in scattering processes. These all have definite quantum numbers: spin $J$, parity $P$, isospin $I$, and in the case of mesons, charge conjugation $C$. The description of their interactions is governed by the general properties of scattering and reaction theory. Guided only by the requirement that interactions should be causal, relativistic and conserve probability, the $S$- (or scattering) matrix obeys analyticity, crossing symmetry and unitarity~\cite{eden,elop}. Almost everything else was modelling. Nevertheless, it was established that the definition of the state in the spectrum of hadrons is a pole of the $S$-matrix in the complex energy plane. This will produce  a textbook \lq\lq resonant'' peak in the cross-section for scattering in a channel to which the resonance couples, if this state is isolated from other resonances and other strongly coupled thresholds. The search for resonances, states in the spectrum of hadrons, is often thought of as bump-hunting, but really it is pole-vaulting. Reconciling unitarity in a  direct channel process with unitarity in the crossed channel gave a whole new meaning to Regge's potential model ideas that angular momentum could be continued from integers to  complex values~\cite{squires-collins}. It was through these $S$-matrix principles combined with Regge theory that the concept of  nuclear democracy and the bootstrap were to be realised.  The power of three of the basic properties, crossing, analyticity and Regge theory, to determine reaction amplitudes was explicitly exhibited in the seemingly magic formula of Veneziano~\cite{veneziano}. This combined poles in energy with zeros in scattering angles: zeros that give hadrons their spin. This remarkable success in representing scattering by a ratio of Gamma functions led to wide consultation of the Bateman manuscript~\cite{bateman} (nowadays something that would be  very much easier with computational resources like {\it Mathematica}). The Veneziano model originally provided a description for $2\to 2$ scattering, see Fig.~1, but this was later generalised to any multiparticle amplitude. This produced a revolution that culminated in the hadron string~\cite{string}. The world sheet of the motion of strings illustrated in Fig.~1  made explicit the bootstrap notion that the sum of all exchanges in one channel generated the hadronic intermediate states in the crossed channel, and {\it vice versa}. The sheet just needed to be pulled in different directions, Fig.~1. Hadrons were oscillating string-like bundles of energy. Their flavour properties resided  at the ends of the strings.  The conservation of probability, embodied in unitarity, which is so important in 
hadron dynamics, was not easy to encode in such approaches. In strong interactions, there seemed no small parameter with which to order the consequences of unitarity. However in nature processes were very readily as strong as they could be.
Nevertheless, the hadronic string picture  enveloped the field for a decade till a microscopic theory took over.  That the flavour properties of hadrons could be re-expressed in terms of quarks (at the end of the strings), led to rules like that of Okubo, Zweig and Iizuka (OZI)~\cite{OZI} that reactions with resonances were dominated by graphs with connected quark lines, Fig.~1. Why was as yet unknown.

\begin{figure}[th]
\centering
\includegraphics[width=14.5cm]{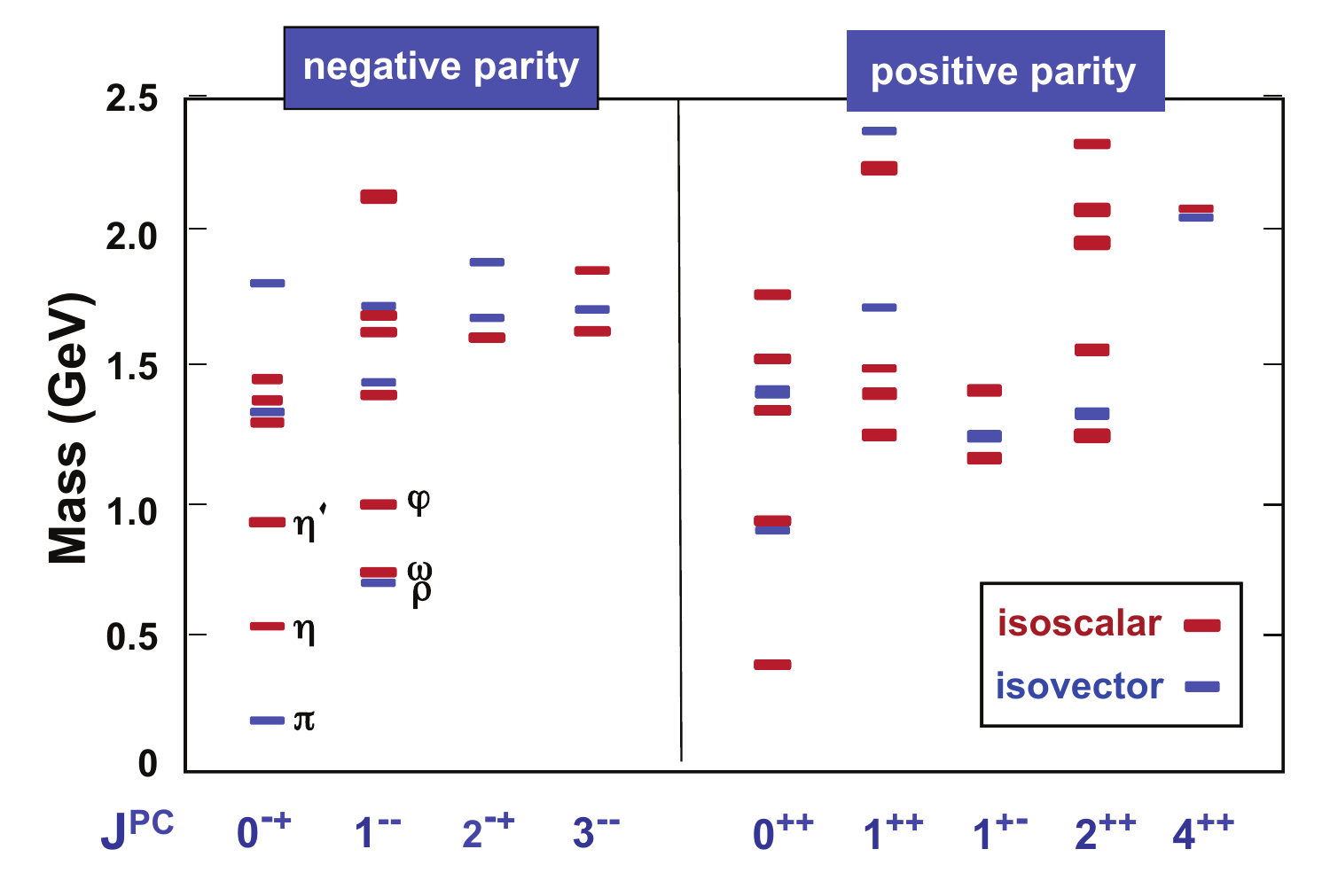}
\caption{The spectrum of $I=0, 1$ light flavoured mesons  revealed by experiment over decades, arranged  in columns of their $J^{PC}$ quantum numbers. The results used to make this plot can be found in the Review of Particle Properties~\cite{pdg}. The members of the ground state pseudoscalar and vector quark model nonets are labelled. }
\label{fig-2}       
\end{figure}
  
Fifty years ago the degrees of freedom in the nuclear world were hadrons, with flavour properties summarised by a quark picture. Since then the number of known hadrons has doubled in both the baryon and meson sectors.  In Fig.~2 is shown the spectrum of lights mesons from the Review of Particle Properties (PDG tables~\cite{pdg}) in columns of their quantum numbers, $J^{PC}$: spin $J$, parity $P$ and charge conjugation $C$. 
 Why is this of interest?  Is this anything more than stamp-collecting, the phrase that Rutherford might have applied to such activity? The spectrum of states of any system is fundamental. It reflects the constituents that make up that system and the forces that bind them together. Nowhere is that more so than for hadrons. The paradigm was of course set earlier by atomic physics. Even if we could not separate an electron from a proton, but only observed the electrically neutral hydrogen atom, we would know from its excitation spectrum that it behaved as if it were built of electrically charged objects bound by the rules of Quantum Electrodynamics (QED). Indeed, the fine structure observed in the atomic spectrum
confirms the  spin-spin and spin-orbit forces inherent in QED.  Can we apply this same reasoning to the inner dynamics of the hadron world?
 Fifty years ago it was known that the ground state baryons, those with spin-parity $({1}/{2})^+$, formed a family with eight members. The isodoublet neutron and proton, the isotriplet $\Sigma$'s in three charges, the isosinglet $\Lambda$, and the isodoublet cascades, $\Xi$. Their properties, masses and flavours, followed from the notion that each was made of three quarks constructed in all possible ways from three flavours of quark, $u$, $d$ and $s$, Fig.~3a. All of these, except for the proton, decay by the weak interaction. If this is switched off, they live forever.
 The discovery of the $\Omega^-$, a state of three strange quarks, completed a decuplet
of  spin-parity $({3}/{2})^-$ ground states was spectacular confirmation of the quark model picture. A similar pattern, shown in Fig.~3b, applied for the light mesons too. This we will discuss in more detail later. 
In the next 10 years this picture extended in  dramatic fashion to {\it charm} and {\it beauty}. Indeed, the November revolution of 1974 proved once and for all that quarks were real. The discovery of the long-lived $J/\psi$ and $\psi'$ 
coupled with a stepwise increase in the ratio of the $e^+e^-$ cross-section to hadrons over that to $\mu^+\mu^-$ proved a new flavour of quark was being excited.  

\begin{figure}[h]
\centering
\includegraphics[width=16.cm]{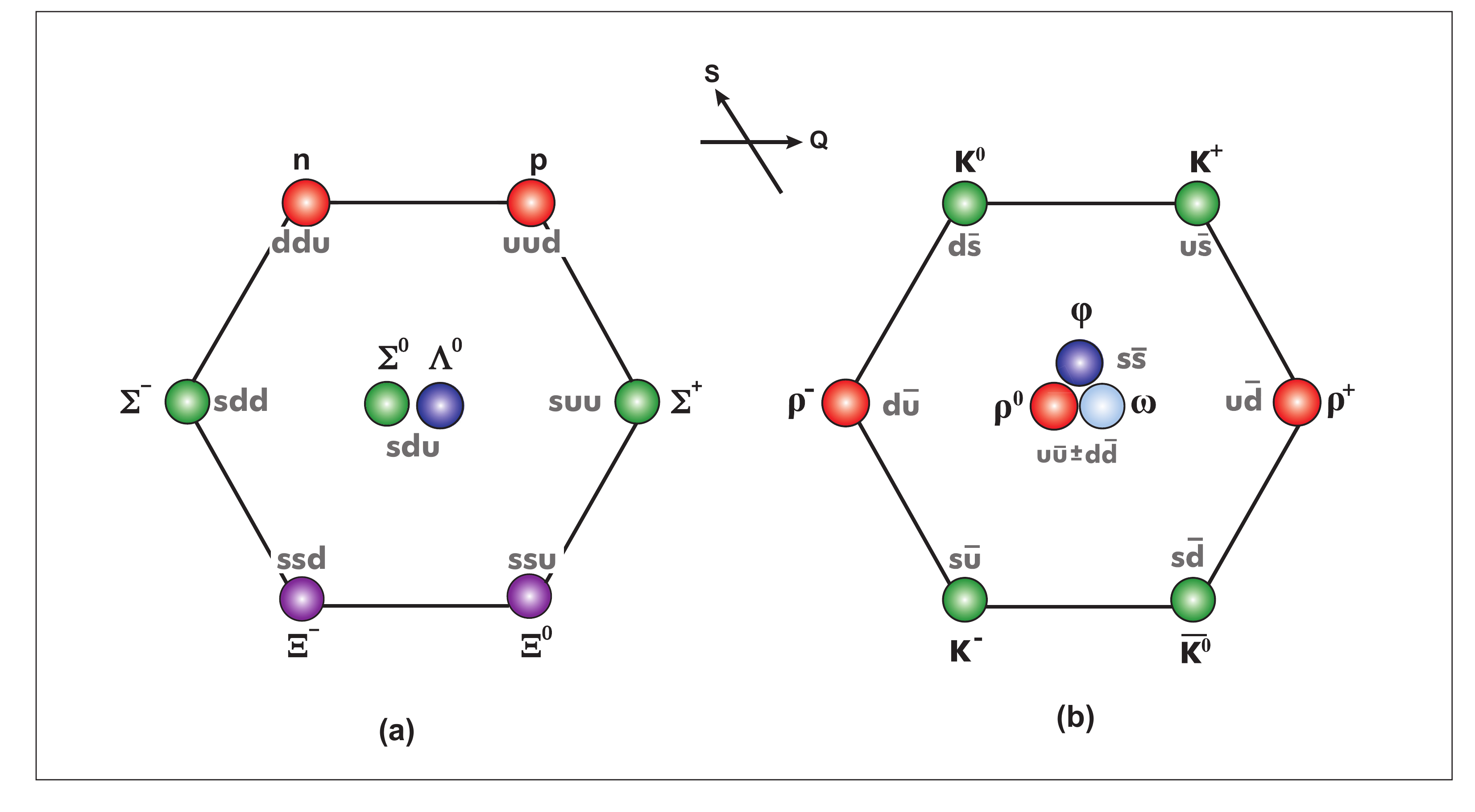}
\caption{(a) The ground state $J^P = 1/2^+$ baryon octet. This includes the nucleons and $\Xi$'s in two charges, the $\Sigma$s in its three charge states, and the neutral $\Lambda$, arranged according to their charge $Q$ and strangeness $S$. (b) The ground state $J^{PC} = 1^{--}$ meson nonet. This includes the three charged $\rho$'s, the  $K*$ and ${\overline K^*}$, and the neutral $\omega$ and $\varphi$, similarly arranged. In each case their simplest composition in terms of $u,d,s$ quarks are shown. The neutral vector mesons have wavefunctions very close to  the idealised $({\overline u}u \pm {\overline d}d)/\sqrt{2}$, ${\overline s}s$. }
\label{fig-3}
\end{figure}

 Earlier, deeper probing than ever before was pioneered by inelastic scattering of electrons at SLAC on hydrogen and nuclear targets, Fig.~4a. There a virtual photon struck hard scattering centres inside the target: hard scattering centres, called partons, with the charges of quarks. Scattering with neutrino beams showed these centres (probed by what we now know is the $W$-boson), have the expected weak charges of quarks too. These quarks carry only 50\% of the proton’s momentum (in the infinite momentum frame). Where is the rest? This was the first indication that the proton contained electrically neutral partons too:  the carriers of the strong force. Now however deeply the proton was probed by the virtual photon imparting higher momentum to the scattering centres, these partons never got knocked out of a region bigger than a few femtometers. Only pions, kaons and protons ever travelled over metres to detectors. Deep inelastic scattering and the spectrum of heavy mesons confirmed that hadrons contained quark degrees of freedom. But these had some strange, almost contradictory properties: they moved freely inside hadrons, but appeared to be trapped.
    It seemed a hadron was like a bag~\cite{johnson-jaffe}, a balloon, perhaps with a fuzzy boundary~\cite{thomas}.
Such bag modelling provided tools for calculating the confining aspects of the underlying theory. Experiment posed challenges for our understanding of hadrons and their interactions: challenges, that would lead to a theory of the strong interaction.

The weak interaction pointed the way. In $\beta$-radioactivity  a neutron, being just 1.3~MeV heavier than a proton, decays by emitting an electron and an electron antineutrino  and turns into a proton in the process $n \to p e^- {\overline \nu}_e$. But what if this was really happening at the quark level:  $d\to u e^- {\overline \nu}_e$. The weak interaction changed quark flavour, just as it changed lepton type. This placed quarks and leptons on the same footing. Forty years ago the theories of the electroweak interactions expressed this in a local gauge theory. The success of this electroweak theory in unifying the weak and electromagnetic interactions, and successfully predicting the existence of the $W$, $Z$ and Higgs bosons, was a major achievement~\cite{weinberg,salam}. This theory was based on the unification of an $SU(2)$ gauge interaction of weak isospin with a $U(1)$ theory of weak hypercharge broken by the Brout-Englert-Higgs mechanism~\cite{brout-englert-higgs}. The most successful theory to date makes predictions for reactions with enormous precision. 
But what of the strong interaction? Was this described by a local gauge theory too? Clearly quarks were the underlying degrees of freedom describing hadrons, but they were surely not knocked out in the deepest inelastic scattering. Quarks could not just be bits of hadrons: bits with fractional electric charge. They had to have their own uniquely defining property. They carried a strong charge we call colour. While quarks have colour, hadrons are colour neutral: in analogy with electrically neutral atoms built of electrically charged constituents. Now the minimum number of quarks in a proton is three. These quarks each carry a different colour. The colour neutral wavefunction is symmetric under permutations of the colours, yet antisymmetric under the interchange of any pair, since quarks are fermions.  I call the colours green, blue and red. You call them blue, green and red. That doesn't alter the colour neutral proton. Naming colours doesn't change the proton. If you now look at the quarks in the  proton through kaleidoscopic glasses and rotate the colours continuously, still the proton doesn't change. This is the basis of the theory that is {\it Quantum Chromodynamics}. Physics is unchanged under local rotations in the colour space. This defines the $SU(3)$ gauge theory that is QCD~\cite{fritzsch-gell-mann-leutwyler}.

\begin{figure}[h]
\centering
\includegraphics[width=15.5cm]{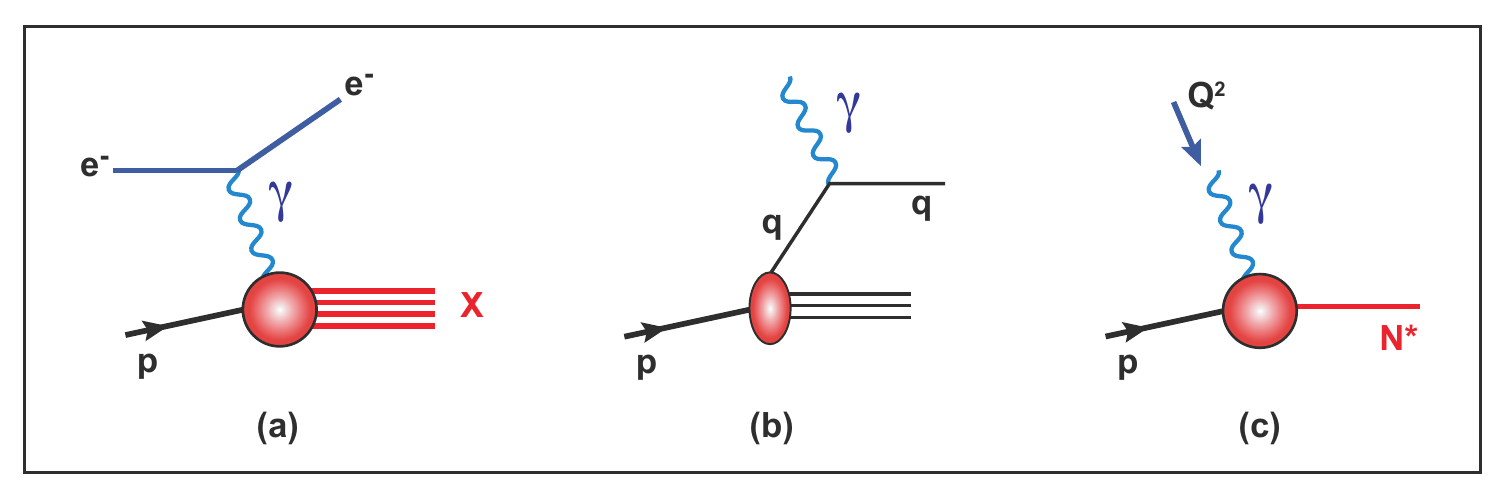}
\caption{Inelastic electron scattering on a proton. The scattered lepton transfers a 4-momentum, $q^\mu$, to the photon that probes the target proton down to a distance $\sim 1/Q$, where $Q^2 = -q^2$. 
(a) depicts deep inelastic scattering (DIS), where the hadron final state is not detected, (b)  shows the lowest order contribution in perturbative QCD to DIS where the photon strikes a single quark $q$ in the proton. The cross-section is then related to the structure functon of the proton. Corrections to this from real and virtual gluon emission lead to the momentum evolution predicted by DGLAP~\cite{dglap}. This is seen to be in very good agreement with experiment in Fig.~8, (c) is the case when the transition to some  exclusive hadronic final state, such as an excited baryon $N^*$, is probed.
}
\label{fig-4}       
\end{figure}

This theory is seemingly simple.  While we know there are three colours, it is useful for later discussion to define the theory for any number, $N_c$. Thus, it contains quark fields in $N_c$ colours for each flavour of quark, with masses $m^f$ depending on flavour, $f$, but not colour. The quark field is a spinor  $\psi^f_i$, which is a vector in colour space with components $i=1,\cdots, N_c$. Invariance under rotations in the colour space forces these quark fields to interact by way of a massless vector boson, the gluons, described by the field $A^\mu_a$ in the adjoint representation of $SU(N_c)$, so $a=1,\cdots, (N_c^2-1)$, through the gauge covariant derivative $D^\mu$. Thus the Lagrangian of QCD is fixed by invariance under colour rotations to be:
\begin{equation}
{\cal L}_{QCD}\;=\; \sum_{f=u,d,s,c,b,t} \bar \psi^{f}_i \left( i \gamma^\mu D_\mu - m^{f} \delta_{ij} \right) \psi^{f}_j\,-\,\frac{1}{4}\, {\cal G}_a^{\mu\nu}\, {\cal G}^a_{\mu\nu}\; .
\end{equation}
The gluon field $A_a^\mu$ forms the field strength tensor ${\cal G}_a^{\mu\nu}\,=\,\partial^\mu A_a^\nu\,-\,\partial^\nu A_a^\mu + g f_{abc} A_b^\mu A_c^\nu$.  The gauge covariant derivative is defined by $D^\mu = \partial^\mu - ig A_a^\mu T_a$,  where the $SU(N_c)$ algebra is specified by the commutation relation $[T_a,T_b] = i f_{abc} T_c$ with  $f_{abc}$ the structure constants of the Lie algebra,
and $g$ is the coupling. Indeed QED is a particular example of such a theory: with a $U(1)$ gauge group. Electric charges commute means $f_{abc} =0$. In QCD, while the kinetic energy of the force carrying gauge bosons is constructed from the field strength tensor as in QED, the required appearance of the $f_{abc}$ term means gluons interact with themselves  with both three and four gluons at one point.
Nevertheless, there is just one coupling specified by $g$. In the Lagrangian these are bare couplings, to go with bare fields and bare masses. But on renormalisation the masses and couplings are defined at some momentum scale, and they \lq\lq run'' with this momentum. 
\begin{figure}
\centering
\includegraphics[width=10.cm]{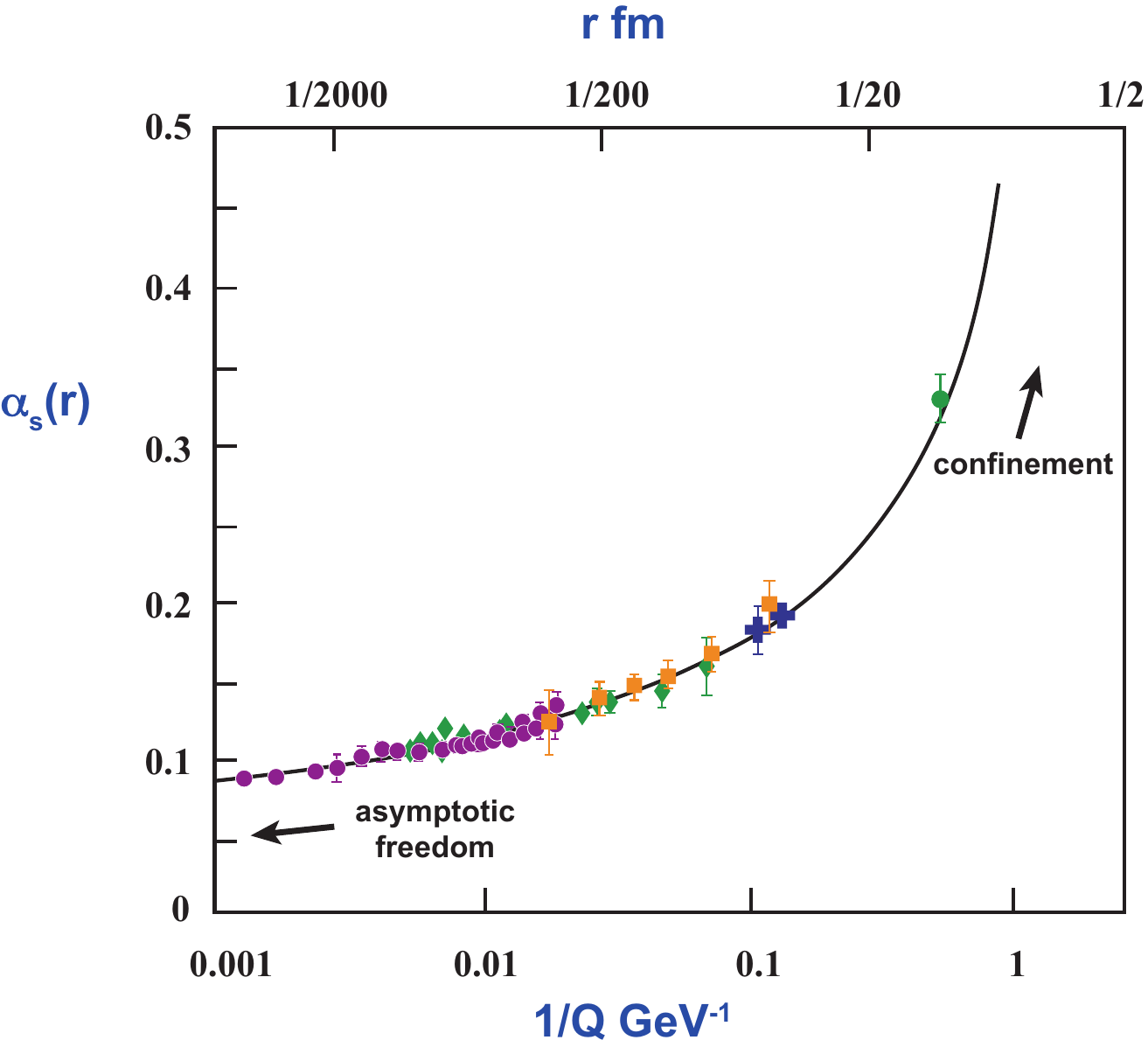}
\caption{The strength of the coupling of QCD, $\alpha_s$, varies with distance $r$, where $r\,=\,1/Q$, with $Q$ the momentum scale that is probed in the interaction, shown on a logarithmic scale. At short distances, the coupling is weaker --- a property known as asymptotic freedom. There perturbation theory applies and this behaviour of the coupling is confirmed by results by data on (a) $\tau$-decays, (b) Lattice QCD, (c) jets in deep inelastic scattering, (d) heavy quarkonia, (e) jet shapes in $e^+e^-$ annihilation, (f) fits to the $Z$-pole, and (g) proton-proton and proton-antiproton production of jets (see \cite{pdg} for references).  At larger distances,  $r > 1$ fermi, the coupling becomes strong. Perturbation theory can no longer be applied. This is the region responsible for colour confinement and the dynamical breakdown of chiral symmetry. The \lq\lq effective'' coupling at distance scales greater than a few fermis may well become constant, and not continue to grow.}
\label{fig-5}
\end{figure}

 From the coupling $g$ one defines $\alpha_s = g^2/4\pi$. This the coupling between quarks and gluons, $\alpha_s(Q^2)$, is a function of momentum $Q$, or equally distance $r=1/Q$. This is where QCD has a truly remarkable property. The coupling runs so that at larger momenta, or shorter distance, it becomes slowly smaller~\cite{politzer-gross-wilczek}. A perturbative view of what is happening deep in the femto-universe applies. This naturally explains why in deep inelastic scattering the struck quark behaves as though it is almost free, barely interacting with the other quarks, except for a slow $Q^2$-dependence. It is asymptotically free, Fig.~4b. This running with distance is confirmed by experiment, as illustrated in Fig.~5. $\alpha_s(Q^2)$ behaves like $1/\ln (Q^2/\Lambda_{QCD}^2)$, where $\Lambda_{QCD}$ is a momentum of 100--200 MeV, that sets the scale at which strong interaction becomes strong. Moreover in $e^+e^-$ collisions at SLAC (the same collisions that discovered particles of hidden {\it charm} and hidden {\it beauty} with the $\psi$ and $\Upsilon$ families) found jets. These were indeed messengers from the parton world. Not only does the size of the $e^+e^-$ cross-section for producing hadrons more or less equal that given by simply summing the squares of the  quark charges in three colours, but the hadrons remember they come from quarks. Grouping the final state hadrons into back-to-back jets, one finds that their angular distribution (relative to the incident leptons) follows that expected of spin-1/2 particles. While two jets dominate, there is a probability proportional to $\alpha_s(Q)$ that a gluon is radiated and a third hard jet emerges to share the total energy of the annihilation. Sometimes this could be clearly separated, in what is called a Mercedes-Benz configuration, Fig.~6. The later $e^+e^-$ collider at DESY found just such events as perturbative QCD predicted~\cite{ellis-gaillard-ross}.
\begin{figure}
\centering
\includegraphics[width=6.5cm]{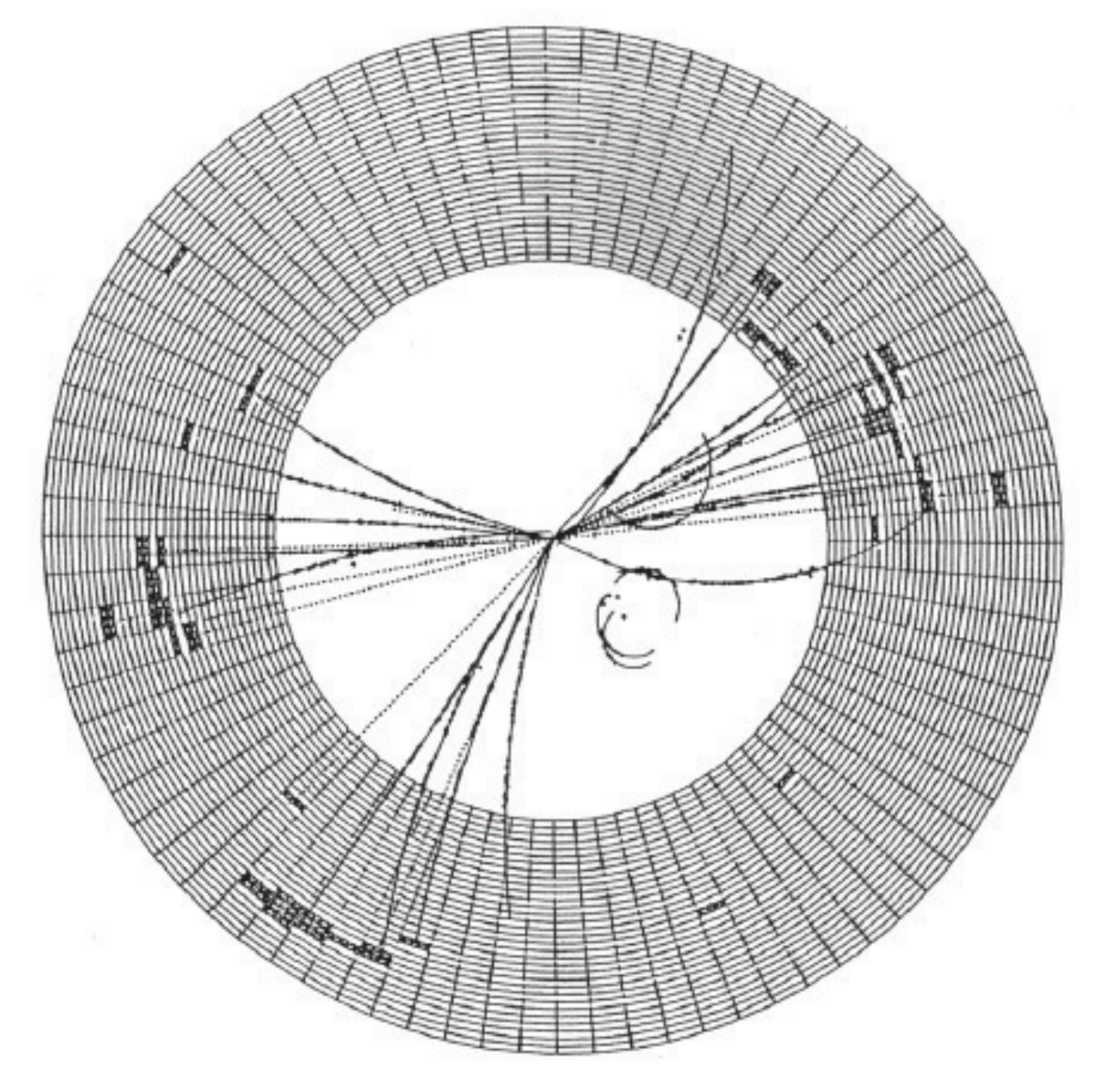}
\caption{Three jet event recorded in the JADE calorimeter at DESY looking along the beam pipe of an electron-positron collision, showing how the hadron tracks cluster in three groups. Detailed analysis of many such events reveals that one of these jets (usually the one carrying the least energy) has a distribution reflecting its origin as a spin-1 gluon. The other more energetic jets are from a quark and an antiquark. }
\label{fig-6}       
\end{figure}
   
While quarks carry colour, the carriers of the strong force, the gluons, carry colour and anticolour. The colour neutral one plays no role in the strong interaction: QCD being an $SU(3)$ gauge theory. Thus there are eight types of gluon. In what is perhaps an apochryphal story, when a famous example of a 3-jet event, probably from TASSO (with a similar example illustrated here in  Fig.~6 by an event recorded by JADE) was presented at a seminar in Oxford claiming that one of these was a gluon jet, Don Perkins scoffed \lq\lq {\it with one event you are claiming eight particles!}''  These messengers from the parton world were studied without worrying about how they ended up as pions and kaons in detectors. The underlying quark and gluon nature of jets was all that mattered.  Not only could QCD be treated perturbatively when leptons annihilate, but in $pp$ collisions too the production of jets of hadrons carrying away large transverse momenta could be calculated in high energy reactions  in good agreement with experiment. The property of asymptotic freedom means many such hard scattering processes can be computed. The success of these predictions meant that we entered almost two decades where the world was perturbative. The fundamental theory had been exposed without the need to access the “dark and messy” mysteries of the strong part of the strong interactions. That could be left for computers to solve, in approaches like that of lattice QCD, or studied in Effective Field Theories.

The current masses of the {\it up} and {\it down} quarks that appear in the renormalised Lagrangian, Eq.~(1), are only a few MeV, very much smaller than the scale set by the running of the coupling, $\Lambda_{QCD}$. It is then a good approximation to the real world to regard these lightest quarks as massless. Then QCD has a chiral symmetry. Helicity becomes a good quantum number, and the left and right-handed spinning worlds decouple. However, in the hadronic universe scalars and pseudoscalars, vectors and axial-vectors are far from degenerate in mass. This breaking of chiral symmetry is one of the most important consequences
of the strong coupling of quarks and gluons, visible in all of low energy hadron physics. However, even this could be rendered perturbative by using a Chiral Effective Theory as proposed by Weinberg~\cite{weinberg-chpt}, and detailed by Gasser and Leutwyler~\cite{chpt}, and a phalanx of later chiral-practors~\cite{chiral-practors}. Though the coupling is strong,  the chiral aspect of QCD for light quarks  means that the interactions of pseudoscalar mesons  are “weak” at low energy and a perturbative expansion in momenta and quark masses can be realised at low enough energies.  $\pi\pi$ and $\pi K$ scattering in the $S$-wave channel accord with this.
However within just a few hundred MeV of threshold,  they do become strong and a wholly perturbative treatment of unitarity proves inadequate. Consequently, various unitarisation procedures have been adopted. These work particularly well for the $P$-waves and show how the near threshold behaviour of these waves both in $\pi\pi$ and $\pi K$ knows about the parameters of the $\rho$ and $K^*(890)$, respecively, just as a dispersive approach would imply.

               An area of hadron physics that appeared to be particularly simple was that of heavy flavours. The spectrum of bound states of charm and anti-charm, and bottom and anti-bottom, were described by a simple interquark potential with two key elements: a short distance Coulomb-like $1/r$ potential generated by one boson exchange plus a confining component that dominates at larger distances that could be represented by a linear rising piece $\propto r$. This Richardson potential~\cite{richardson}, shown in Fig.~7a, had to have relativistic, spin-spin and spin-orbit corrections~\cite{eichten}, but this was well developed from atomic physics.
\begin{figure}
\centering
\includegraphics[width=14.cm]{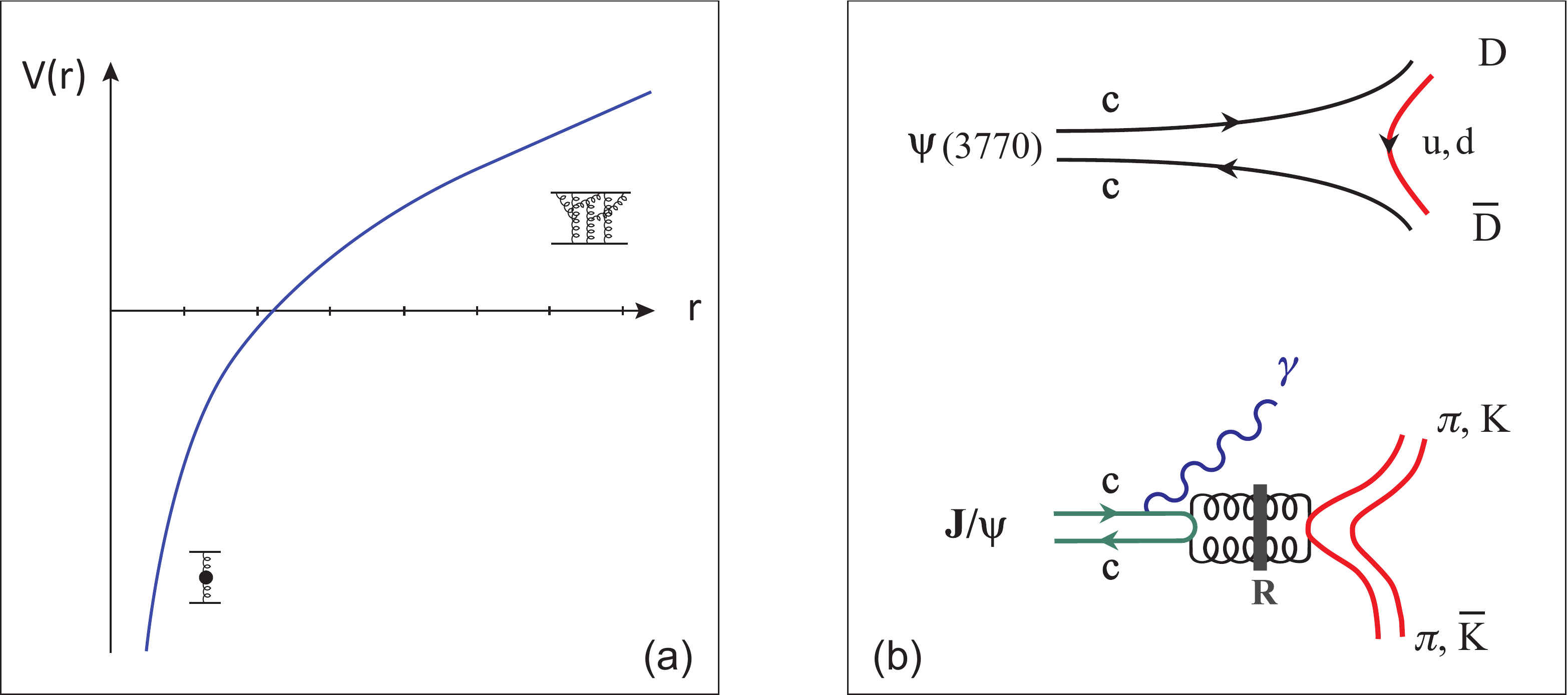}
\caption{(a) Interquark potential as a function of the distance apart of the heavy quark and antiquark. At short distance this is given by one gluon exchange as required by asymptotic freedom. At distances greater than a few fermi it has a linear confining component controlled by multi-gluon exchanges. This potential~\cite{richardson} is a good approximation for both charm and bottom quark systems.
Of course, between light quarks the picture of a non-relativistic potential makes little sense. (b)  quark line diagram for a heavy charmonium state, like the $\psi(3770)$ to decay into charmed mesons, ${\overline D}D$. For the $J/\psi$ such channels are not accessible and the ${\overline c}c$ quarks have to annihilate for it to decay by first producing gluons. When this is a radiative decay,
the  two gluons can have $2^{++}$ quantum numbers, which may resonate, producing a state $R$ that then decays to $\pi\pi$ or ${\overline K}K$, as with the $\xi(2230)$ claimed by BES~\cite{jinshan}. }
\label{fig-7}       
\end{figure}
 These hidden heavy flavour states, particularly those with the same quantum number as the photon, could be studied in the $B$-factories built at SLAC and KEK with their BaBar and Belle detectors, respectively. Their aim was primarily to measure the weak matrix elements precisely and test, together with the $\phi$-factory DAPHNE at Frascati, whether the {\it CP} violation in the Standard Model was sufficient to understand the matter-antimatter asymmetry of the Universe. The electroweak interaction changes flavours: the heavy leptons decay into lighter leptons, and quarks of heavy flavour into the light $u$, $d$, and $s$ flavours. While the mass eigenstates are what experience the strong force, the $W$-boson sees rotations of these: multi-dimensional rotations parameterised by the CKM matrix~\cite{CKM}.  The $B$-factories aimed to measure these matrix elements with precision. The philosophy was that only fundamental parameters of the Standard Model mattered: no need for the messy hadronic bits. To test {\it CP} violation, one wanted to compare, for instance, $B$ decay to ${\overline D} K$ with $B$ decay to $D {\overline K}$. However, the $D$ mesons are unstable and both decay chains readily lead to a common final state, like $\pi\pi K{\overline K}$. That is good as the interference between these decay paths can magnify the tiny {\it CP} violating effects. But, and there is big {\it but}, one needs to understand how these are influenced by the different strong final state interactions in $\pi\pi K$ and $\pi\pi {\overline K}$, and then each with ${\overline K}$ and a $K$ respectively. So while the aim was to measure supposedly perturbative electroweak parameters generated by the emission of a $W$-boson over a time of  $10^{-25}$ s., this short distance interaction is seen through a haze of strong interactions over distances and times 100 times longer.   Strong interactions provide the biggest uncertainty on the determination of the  CKM matrix parameters, and so incertitude in the closure of its unitarity triangle. This realisation has changed the mode of addressing what was naively thought to be fundamental,  perturbatively describable, interactions. Any initial or final state that involves hadrons requires the challenges of strong physics to be confronted. The years of being brushed away into dark corners would soon be over.

\newpage
\section{Current quarks and perturbation theory}               

We have so far glibly talked of \lq\lq quarks'' that are the constituents of a proton or of the $J/\psi$, quarks in jets of hadrons produced by $e^+e^-$ annihilation, and the quarks probed in deeper inelastic scattering, as though these were the same, which they are not. When the proton is at rest, it can be thought of as made of three constituent quarks, two {\it up}s and a {\it down}. These have masses of about 300-350 MeV, a third of the mass of a proton. They are the objects that underlie the quark model of Fig.~3~\cite{isgur-karl,capstick-isgur}. Now consider this proton moving essentially at the speed of light. It becomes flattened like a pancake in the direction of motion. Its internal constituents, {\it partons}, which are in fact quarks, antiquarks and gluons, share the now large momentum of the proton, carrying a fraction $x$ each. The cross-section in deep inelastic scattering measures (in the case of electron scattering) the probability of the photon interacting with the electrically charged partons as a function of $x$. What the photon \lq\lq sees'' changes with its  wavelength $~1/Q$.  When this is $1/3$ to $1/10$ of the size of a proton, the parton distribution is dominated by {\it valence} quarks that, like constituent quarks, carry the distinctive flavour properties of the proton. In total these carry less than 50\% of the proton's longitudinal momentum. Indeed, the valence distribution peaks around $x\sim 0.15$.
 While in spectroscopy the constituent quark is a distinct object, one can imagine that in deep inelastic scattering one is looking inside this object and finding that really it is quark, say an {\it up} quark, surrounded by a cloud of gluons and a small sea of ${\overline q} q$ pairs, all sharing the proton's momentum. As $Q$ increases, the probability that the photon strikes a quark in the sea increases, and in turn this impulse causes more gluons to be radiated, which increases both the density of gluons and of the sea. In such very deep inelastic scattering is the nearest we get to probing barely dressed quarks, the current quarks of the renormalised Lagrangian Eq.~(1).

The asymptotic freedom of QCD allows scattering with large momentum scales to reveal the quarks almost naked. One of the most spectacular demonstrations of this  is the application of perturbation theory success to the evolution of nucleon structure functions over a huge range of momenta, probed by photons in electron scattering, and by $W$’s in neutrino interactions. 
As is well known the structure function, $F_2^p$, shown in Fig.~8 reflects the dependence of the underlying parton distributions (quarks, antiquarks and gluons) on the fraction of the proton's momentum, $x$, carried by each parton in the infinite momentum frame.   How these change with the momentum $Q$ of the probing photon, Fig.~3b, is described by perturbative QCD at some appropriate order. The Dokhsitzer-Gribov-Lipatov-Altarelli-Parisi (DGLAP) evolution~\cite{dglap} takes into account gluon emissions by the struck parton. This is all that is required according to the property of asymptotic freedom. Correlations between partons are suppressed at large probing momenta by powers of $1/Q^2$. That DGLAP formulation describes the momentum evolution from SLAC energies, where experiments first established the parton model, to those of HERA is impressive. Indeed this now extends into the LHC domain, where parton distributions determine the cross-sections that are observed for $WW$ production for instance. With suitable target mass corrections~\cite{accardi}, in turn these can be linked to the precision results of electron scattering in the valence regime accessed at Jefferson Lab. DGLAP  rules ok~\cite{dglap}, as seen in Fig.~8. This is not only true for unpolarised structure functions, but for their polarised extensions too~\cite{nobuo}. 
\begin{figure}
\centering
\includegraphics[width=12.5cm]{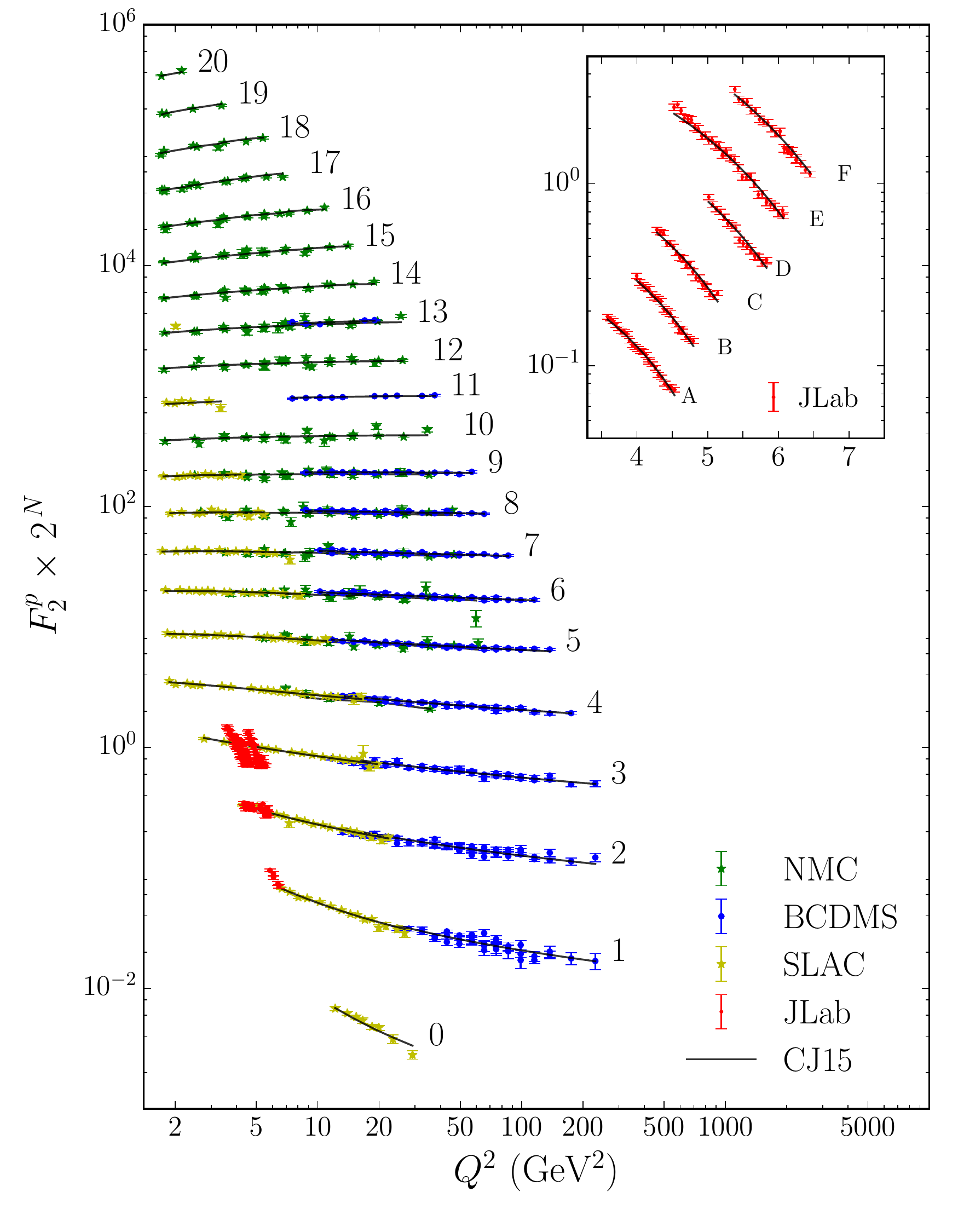}
\caption{The structure function $F_2(x, Q^2)$ of the proton, plotted as a function of $Q^2$ and multiplied by a factors of $2^N$ where $N$ labels the lines in the main plot. To enable the data to be displayed on one plot, a different $N$ is used for each $x$-value) $N=0$ $x=0.85$, 1) 0.74, 2) 0.65, 3) 0.55, 4) 0.45, 5) 0.34, 6) 0.28, 7) 0.23, 8) 0.18, 9) 0.14. 10) 0.11, 11) 0.10, 12) 0.09, 13) 0.07, 14) 0.05, 15) 0.04, 16) 0,026, 17) 0.018, 18) 0.013, 19) 0.008, 20) 0.005.  The insert with JLab data show the $Q^2$ evolution over the range accessible at JLab at 6 GeV. The data and curves are at fixed scattering angle $\theta$ with A) $38^o$, $N=0$, B) $41^o$, $N=1$, C) $45^o$, $N=2$, D) $55^o$, $N=3$, E) $60^o$, $N=4$, F) $70^o$, $N=5$. The lines are from the CJ15 fit~\cite{accardi} with DGLAP evolution at next-to-leading order. In that same paper can be found the references to the experimental data shown.}
\label{fig-8}       
\vspace{-0.2cm}
\end{figure}

\begin{figure}
\centering
\includegraphics[width=9.cm]{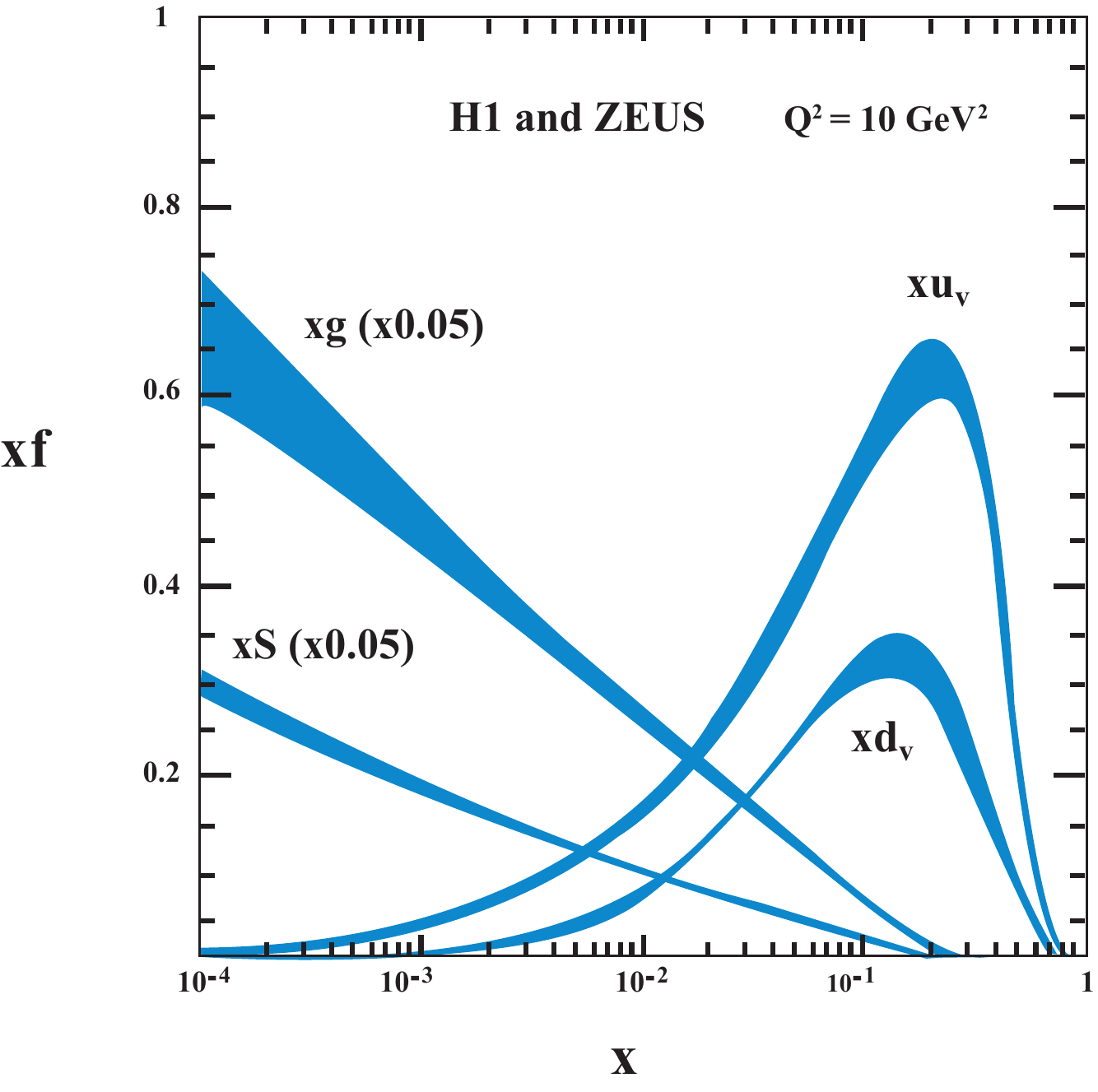}
\caption{Representation of the Parton Distribution Functions, $xf(x)$, from the HERAPDF 2.0 analysis~\cite{amanda} for the valence quark $u_v$, $d_v$, sea $S$ and gluon at $Q^2 = 10$ GeV$^2$, the sea and gluon distributions being scaled by a factor of 1/20.
The valence distributions are defined to be $q_v = q - {\overline q}$ for the {\it u} and {\it d} quarks. The sea is given by $S = {\overline u} + {\overline d} + s + {\overline s}$. These PDFs describe the probability of finding a particular parton inside the proton carrying a fraction $x$ of the proton's longitudinal momentum along the light cone when probed at the given value of $Q^2$, here 10 GeV$^2$. These accord with the evolution with $Q^2$ given by the NLO pQCD fit CJ15~\cite{accardi} shown in Fig.~8.}
\label{fig-9}       
\end{figure}

While perturbative QCD~\cite{dglap} describes the $Q^2$ dependence of the structure function of the proton, $F_2(x, Q^2)$,  remarkably well, as seen in Fig.~8, its form at any particular $Q^2$ is an intrinsic property of the proton, not given by perturbation theory. Underlying such a structure function are the PDFs, the parton distribution functions, which are the probability of finding a particular parton inside the proton carrying a fraction $x$ of the proton's longitudinal momentum. These have been studied over decades, for instance~\cite{cteq,mrst}.  $x$ times these distributions are shown in Fig.~9 at $Q^2 =10$ GeV$^2$ from a combined analysis of H1 and ZEUS data from HERA~\cite{amanda}. The data used to determine these distributions appear as part of  Fig.~8. Since the PDFs are probabilities, the integral of the sum of these is, in principle, the number of partons, which is not a defined quantity. There are infinitely many soft gluons inside a proton.  However, weighted by $x$ as in Fig.~9, the sum is the total fraction of the proton's momentum carried by partons, which is of course 1. At larger $x$ the distributions are dominated by valence quarks, the difference of quark and antiquark, that gives the proton its characteristic flavour. Integrating $u_v$ and $d_v$ gives 2 and 1, respectively. Away from the valence regime, at smaller $x$, emphasised by Fig.~9 with its logarithmic scale in $x$, gluons and the sea (which sums the ${\overline u}$, ${\overline d}$ with $s + {\overline s}$ contributions) dominate enormously.
At very small $x$, which may be accessed at a future Electron-Ion Collider, the density of gluons may be so great that it has been speculated that a new state of matter, a \lq\lq colour-glass'' condensate of gluons~\cite{color-glass}, may appear. 

\begin{figure}
\centering
\includegraphics[width=6.5cm]{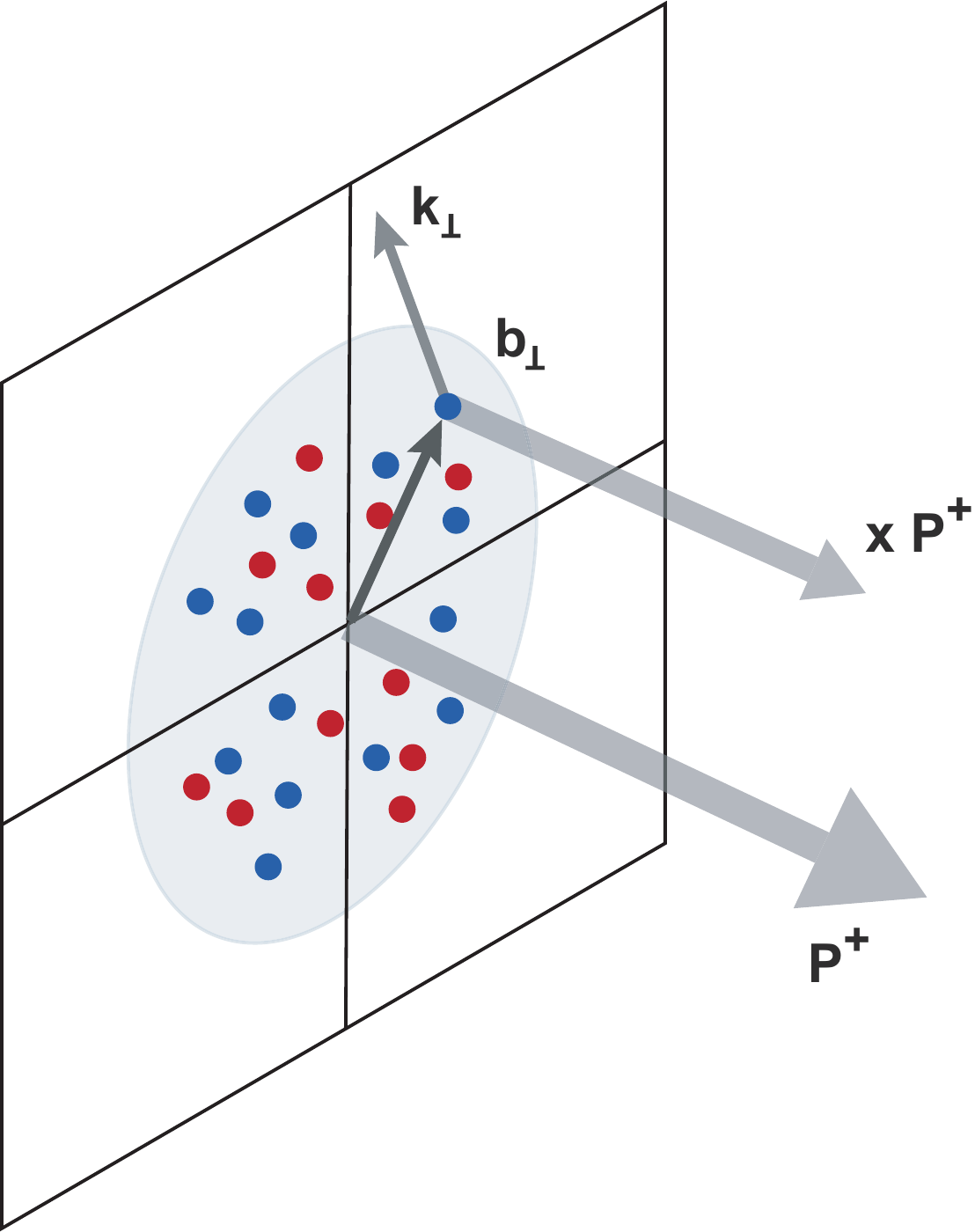}
\vspace{2mm}
\caption{A proton moves along the light-cone with momentum $P^+$. To an on-coming photon it appears like a pancake. When the wavelength of the photon is short, it will strike just one of the proton's components or {\it partons} that is carrying  a fraction $x$ of the proton's longitudinal momentum. This struck parton will have a transverse momentum ${\bf k_\perp}$, by virtue of being confined inside the proton. Its transverse position will be given by a vector ${\bf b_\perp}$ in the transverse plane.  }
\label{fig-10}       
\end{figure}

On the way we have learnt that the deep inelastic scattering (DIS) on a nucleus is not  merely scattering on a collection of free nucleons.
The common form of iron, $^{56}Fe$ for example, with 26 protons and 30 neutrons yields PDFs that are not just that of a sum of the individual nucleons. Indeed the results first found by the European Muon Collaboration (EMC~\cite{emc} and explored further by NMC~\cite{nmc})  reflect not just the expected Fermi motion  but point to key correlations between nucleons in each nucleus~\cite{arrington}. A property that is currently under study.

Electromagnetic formfactors of the nucleon are, like parton dsitributions, one of its intrinsic properties. However, these are related. The parton structure is in fact described by a multi-dimensional distribution, a Wigner function. When the proton is travelling along the light-cone, as in Fig.~10, the Wigner function depends  not only on the longitudinal momentum fraction, $x$, of the struck parton, but also on its transverse momentum $k_\perp$ and its impact parameter $b_\perp$, both of which are two-dimensional vectors, Fig.~10. Integrating the Wigner function over these transverse dependences gives the familiar one dimensional parton distributions, Fig.~9, whereas the electromagnetic formfactors reflect  the transverse aspects of these distributions. Integrating the Wigner function over impact parameter gives Transverse Momentum-dependent Distributions (TMDs) that are being studied in Semi-Inclusive Deep Inelastic Scattering, where the struck parton in Fig.~3b, materialises as a 
particular detected hadron, for instance a pion. Complementary are the integrals over transverse momentum, which give Generalised Parton Distrbutions (GPDs) that are explored in Deeply Virtual Compton Scattering (DVCS) and Meson Production (DVMP), see Fig.11.   

\begin{figure}[t]
\centering
\includegraphics[width=14.cm]{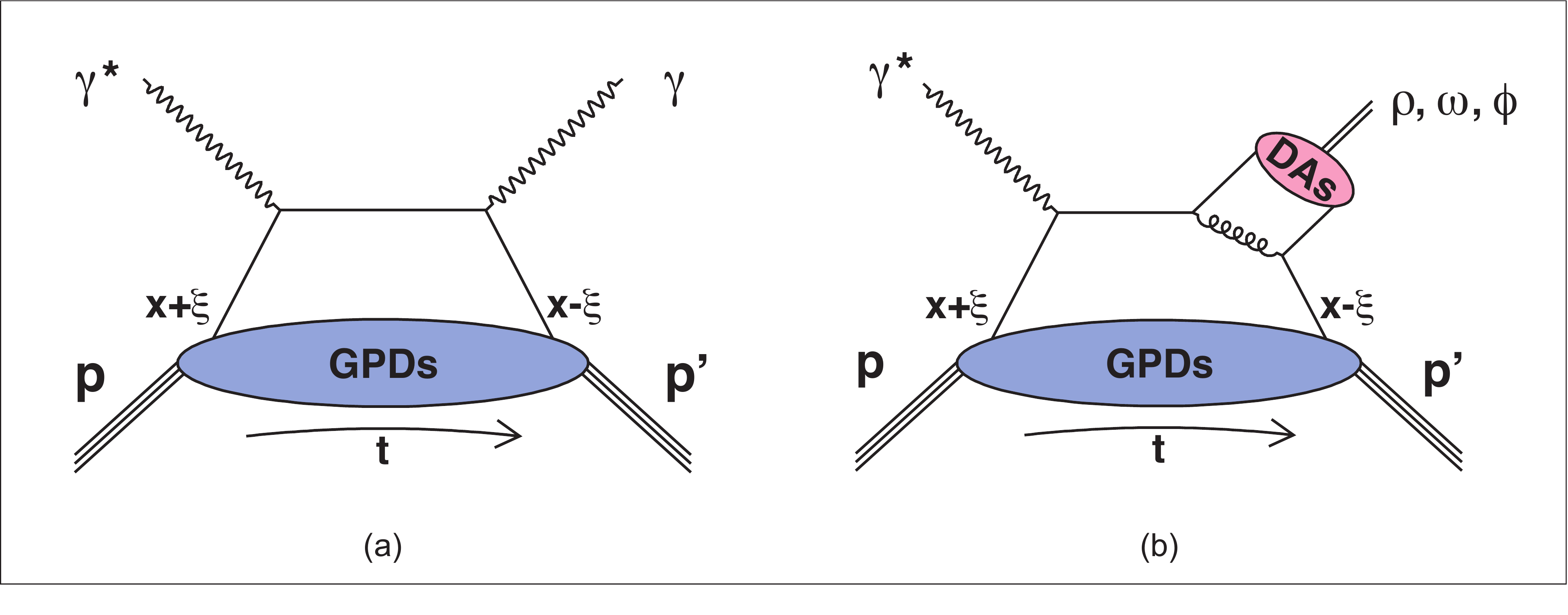}
\caption{The \lq\lq handbag'' diagrams representing Deeply Virtual Compton Scattering (DVCS) and Deeply Virtual Meson Production (DVMP). As in Fig.~3, the initial (virtual) photon is produced by a scattered electron. $x\pm \xi$ are the fractions of the proton's momentum along the light-cone carried by the {\it active} parton before and after the collision. }
\label{fig-11}       
\end{figure}

40 years ago we might have thought that the valence quarks in the proton not only carry the flavour of the proton (its baryon number and charge), but also its momentum and the spin.
We first learnt the valence quarks carry less than 50\% of its longitudinal momentum at $Q \sim 3$ GeV and then more recently that they contribute only 30\% of the proton's spin $1/2$. Where is the rest? in gluons or in the orbital motion of the quarks? 
 The proton is a swirling universe of partons and the way the spin sumrule~\cite{ji-spin} is fulfilled has been a matter of intense debate:
\be
\frac{1}{2}\;=\; \frac{1}{2}\,\Delta q\,+\,L_q\,+\,J_g\; ,
\ee
where $\Delta q$ is the contribution of the spin of the quarks, determined by their polarised parton distributions, integrating over all $x$. Similarly,  $L_q$ is the contribution from the orbital motion of the quarks and $J_g$ the total contribution from gluons. In different formulations $J_g$ can be divided into the spin component of the gluons and their orbital motion, but how (and if) these definitions are related to experimental observables is being resolved~\cite{lorce-leader}.

\begin{figure}[b]
\includegraphics[width=14.cm]{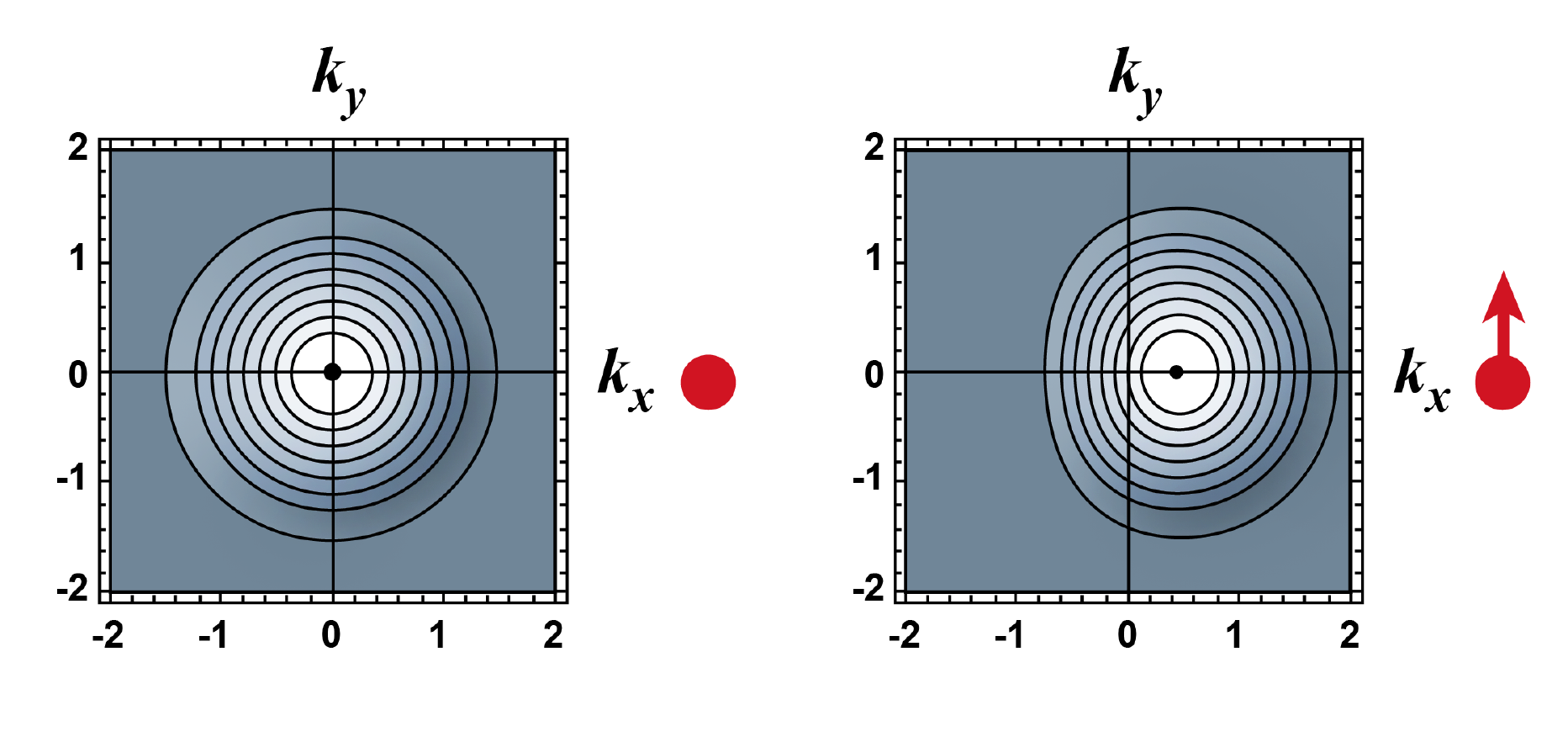}
\caption{A contour plot of  the $k_\perp$ distribution of unpolarised quarks inside a  proton that is coming towards you. On the left is for an unpolarised proton: the parton distribution is symmetric.
On the right is shown how this changes if the proton is polarised in the $y$-direction, indicated by the arrow on the right. One sees the quark distribution distorted and shifted. This change contributes to what is called the Sivers function. Indeed, experiment reveals that when such distributions are decomposed into flavours, the {\it up} quarks shift one way and the {\it down} the other way~\cite{prokudin-musch}. These effects allow the angular momentum carried by quarks to be determined.}
\label{fig-12}       
\end{figure}

 To understand orbital motion whether of quarks or gluons requires a connection between the momentum of each parton and its position, Fig.~10.  We move from the one dimensional momentum distribution of partons (Fig.~9) to their multi-dimensional generalisations. Transverse Momentum-dependent Distributions (TMDs) describe the correlation of the polarisation of the quarks of different flavours with the polarisation of the proton.
By measuring different  observables  in Semi-Inclusive Deep Inelastic Scattering (SIDIS), TMDs combined with  fragmentation functions (as a quark materialises as a detected hadron) can be determined. The feasibility of separating TMDs~\cite{bacchetta,boglione} has been shown using results from Hermes~\cite{hermes,boglione} and COMPASS~\cite{compass-structure,bacchetta} providing a multi-dimensional image of the proton. An illustrative example is provided by the Sivers function, Fig.~12.   For an unpolarised proton, moving along the $z$-direction, the ${\bf k_\perp}$ distribution of unpolarised quarks is naturally symmetric, as shown in the contour plot of Fig.~12 left. If the proton is polarised in the $y$-direction, as in Fig.~12 right, the distribution of unpolarised quarks becomes distorted. It shifts in the $x$-direction, naturally generating an orbital motion of the quarks.  Present studies indicate $u$ quarks move one way, $d$ the opposite way. These contribute cancelling contributions to the total orbital angular momentum of the quarks. Of course, to satisfy the spin sum-rule, Eq.~(2), in every interaction requires correlations of one parton with another. Such correlations are naturally implied by colour confinement. How this works is a challenge for the next decade.

SIDIS studies will open new insights into the workings of QCD.   The formalism for extracting the complementary Generalised Parton Distributions was laid out more than 25 years by M\"uller {\it et al.}~\cite{mueller}, and then by Ji~\cite{ji-gpd} and  by Radyushkin~\cite{radyushkin} from  measurements  of Deeply Virtual Compton Scattering and Meson Production,  Fig.~11. A program for a decade has been set out at Jefferson Lab to test and then utilise the factorisation theorems of QCD to map out such distributions with precision. 
The next decade is likely to provide a whole new picture of the workings of the proton and its partonic structure. JLab at 12 GeV will explore the regime dominated by  valence quarks. The sea of ${\overline q}q$ pairs and clouds of gluons will have to wait for a future Electron Ion Collider~\cite{eic-whitepaper}. There scattering on  polarised light nuclei, such as  deuterium or $^3He$ with spectator tagging~\cite{weiss-tagging}, will allow the neutron to be uniquely explored as well as the proton. Multi-dimensional imaging is likely to have as profound an impact on our understanding of the workings of the nucleon as tomography has had for the human body~\cite{radici}.

\newpage  
\section{Hadrons in QCD: resonances, glueballs, hybrids}

If the definition of a hadron in QCD is that it is  colour neutral, then ${\overline q}q$ mesons and $qqq$ baryons  
are the simplest structures. However, there are of course  many other combinations that are can be colour singlets: glueballs of minimal $gg$ or $ggg$
components, or hybrids mesons and baryons of ${\overline q} gq$ or $qqgq$, or multi-quark states of ${\overline{qq}} qq$ or $q{\overline q}qqq$, {\it etc.}
Forty years ago the search for these began in earnest. States appearing in ${\overline p}p$ annihilation were thought to be where tetraquark systems in their different forms would be readily detected. They were called \lq\lq baryonium''.
Colour neutral diquark-antidiquark systems could emerge from  $\,6\otimes{\bar 6}\,$ or $\,{\bar 3}\otimes 3\,$  of colour. Some were expected to be long-lived. Narrow states
were found, and generated much excitement. Almost all proved to be anomalies of small statistics and disappeared as more data were taken. Much the same happened later with the first pentaquark state, the $\Theta^+(1540)$,  claimed in the $K^+n$ mass spectrum~\cite{Theta}. A repeat experiment with much higher statistics found no significant evidence for this~\cite{DeVita:2006aaq}. 

The search for glueballs first concentrated on charmonium ${\overline c}c$ decays. The long lifetime of the $J/\psi$, for instance, was explained by  asymptotic freedom, Fig.~7b. The charmed quarks had to annihilate, as the $J/\psi$ was well below the open charm ${\overline D}D$ threshold. With a mass of 3~GeV,
the process was thought perturbative and annihilation through at least three gluons was required (to conserve parity and charge conjugation). In radiative $J/\psi$ decays, only two gluons are required, Fig.~7b. These gluons
would then create lighter quarks, which in turn materialised mainly as combinations of pions or kaons. Charmonium decays should thus be glue-rich and it is there that glueball signals should be found.
Indeed, similar dynamics might be sought in double $\phi$ production in $\pi N \to \phi\phi N$, with the $\phi$'s observed in their ${\overline K}K$ decay modes.
Tensor glueballs, indeed a whole series: $g_T(2050),\, g_{T}^\prime(2300),\, g_{T}^{\prime\prime}(2350)$, were claimed in the $\phi\phi$ spectra observed at BNL~\cite{lindenbaum}, and the $\xi(2230)$ in $J/\psi\to\gamma{\overline K}K$ decay~\cite{jinshan}.
All these states have disappeared as data of higher statistics have been produced, or the reasons to think they were really glueballs have vanished. 

Models had suggested the lightest glueball would be a scalar. Among the scalars, even discounting the nine lightest, there were still too many isosinglet $f_0$'s, Fig.~2, for another quark model nonet: the $f_0(1370)$ (if it exists), the $f_0(1510)$, the $f_0(1720)$, for instance. Was one of these a glueball, or had a large glue admixture?  Historically the very lightest scalar, the $\sigma$,  was introduced to represent the second longest range nuclear force, the correlation of two pion exchange, and seen as a broad enhancement in $S$-wave $\pi\pi$ interactions. It was for long not clear that this really was a resonance, and it dropped out of the PDG tables. However, it strongly reappeared when better tools of analytic continuation robustly established its underlying pole $f_0(500)$~\cite{colangelo,pelaez-sigma}. Some have suggested this state has a large glue component~\cite{minkowski-ochs}.

\subsection{Decays: the essence of a hadron's life}
We have learnt that the structure of a hadron  depends on the distance scale at which it is probed and in what frame of reference. When at rest, or in relatively slow motion, we have known for more than 40 years that many of the properties of the ground state baryons, Fig.~3a, can be understood in terms of just three constituent quarks. However, even the proton is continually turning into a neutron and a positive pion, by virtue of the
uncertainty principle, or into a $\Delta^{++} \pi^-$ too,  depending on whether a $\,{\overline d}d\,$ or $\,{\overline u}u\,$ are created. This is happening every $10^{-23}$ second. Indeed in deep inelastic scattering, the virtual photon emitted in the electron interaction will sometimes strike a quark in the pion cloud. The proton regularly is four quarks and an antiquark. 

For excited baryons, the spectrum, akin to that of mesons shown in Fig.~2, is categorised by $J^P$. Forty years ago knowledge of this relied almost entirely on the partial wave analysis of data on $\pi N$ scattering. Over the past 15 years this has been supplemented by large datasets on photoproduction with polarised beams on polarised targets and in some cases measurements of  the polarisation of the final states~\cite{beck,thoma}. While baryons may contain diquark clustering as an effective degree of freedom, these are definitely not pointlike. Consequently such a simple modelling as an explanation  of so called {\it missing states} can be discounted. Indeed, the new precision data
on polarisation asymmetries has not just tidied up the spectrum below 2 GeV by filling in some of the {\it missing states}, it has brought greater certainty  to a remarkable extent~\cite{pdg,anl-osaka,bonn-gatchina,mrp-baryons}.  All these excited states decay at least a  million, if not a billion times, more quickly than the ground states. For instance, the $N^*(1520)$ decays to $N\pi$ and $N\pi\pi$. In traditional nuclear theory you might think of this as just a materialisation of the pion cloud.

In principle deep inelastic scattering (DIS) on an $N^*$ would determine its parton distribution.  Of course, excited baryons do not live long enough for such a probing to be recorded. However, one can study exclusive scattering as in Fig.~4c, where a proton makes a transition to an $N^*$. Of course, the particular $N^*$ has to be extracted from a multihadron final state, and this may have its own uncertainties. Nevertheless, such transitions for a range of $Q^2$ have been under experimental study  over the past decade~\cite{burkert,gothe,mokeev}, with more data to come in the next. These will hopefully provide information about the structure of some of the dominant
excited baryons to be compared with modellings of strong coupling QCD, which we mention later.

The creation of additional ${\overline q}q$ pairs is the very substance of hadron decays~\cite{barnes,isgur-geiger,capstick-roberts,pennington2}.
If we consider  the nonet of vector mesons: $\rho$, $K^*(890)$, $\omega$ and $\phi$, Fig.~3b.
We know that this is \lq\lq ideally mixed'' with distinct hidden flavours for the neutral members
because of their decays and mass splittings. The $\omega$ with a mass that is degenerate with the $\rho$
indicating it too is made of $u$ and $d$ quarks. The $\rho,\, \omega$ are 120 MeV lighter than the $K^*$ with its one strange quark,
and the $K^*$ is 120 MeV lighter than the $\phi$ with its ideal ${\overline s}s$ composition.
Each state decays by creating a ${\overline u}u$ or ${\overline d}d$ pair, making the $\phi$ naturally decay to ${\overline K}K$.
This pattern equally applies to the tensor mesons, and all the states we have found with the maximum spin for a given mass. 

Such mesons, $M$, are readily created in peripheral production processes like $\pi^\pm p\to M N$, (which nucleon $N=n,p$ balances the electric charge depending on the charge of the resonant state $M$ that is produced) that were much studied in the '70s and '80s. Their results painted a picture in which the resonance $M$ is created and then at some subsequent time it decays into lighter mesons. This picture underlies the simple isobar model, much used in analyses of experimental data over the past 40 years. The reason this picture \lq\lq works'' and provided 
evidence supporting  a simple ${\overline q}q$ sructure of these mesons, 
is because all the vector meson decays are $P$-wave interactions that {\it barely} disturb the ${\overline q}q$ components.  The ${\overline q}q$
component of their Fock space dominates. That, of course, does not mean that they do not have multiquark and colour singlet meson components too. These are just less important, as indicated by Fig.~13.
Indeed all mesons with the maximum spin for a given mass, those said to lie along the leading Regge trajectory, all appear to be dominated by their ${\overline q}q$ components. Since these states are the most easily identified in hadroproduction processes, these have shaped our expectations of ideally mixed quark model multiplets. However, we are learning that this may not be the natural order of things. The lower lying states, such as  the scalars and axial vectors, appear differently.

The degrees of freedom of QCD are of course quarks and gluons. Inside a hadron there is a whole universe of these. In  building hadrons the ${\overline q}q$ pairs and gluons cluster round a single quark to make the effective degree of freedom we recognise as a \lq\lq constituent quark''. In decays new dressed quarks are formed that create the final state particles. The effective degrees of freedom are then  not just coloured quarks and gluons.  Forces between coloured quarks and colour neutral pre-hadrons coexist. While it is seductive to think of a hadron, like the $\rho$ or $a_1$,  being created in some reaction and then at some later time decaying, this picture does not reflect the fact that as such a meson resonance is being created, components other than just one ${\overline q}q$ pair must be forming for these states to decay into $\pi\pi$ and $3\pi$, respectively, all within the same femto-universe. The coupling to nearby hadronic channels, particular those with $S$-wave couplings, affect the mass of the state. Indeed, when such couplings are particularly strong they may force the hadron resonance to be very close to this threshold, and the multi-hadron components may dominate its Fock space. Where this happens the state may be thought to be dynamically generated by interhadron forces, just as the bootstrap had presumed~\cite{chew}. However, it could well be that the presence of quark model {\it seeds} is critical in driving this.  While exchange forces failed to bootstrap the majority of  resonance poles  in the dispersive approach to hadron dynamics, the presence of intrinsic states that the quark model naturally embodies proved to be essential, as foreshadowed by the discussion in dispersion theory about the role and presence of CDD poles~\cite{cdd}. Now 40 years later when the supremacy of quark dynamics appears totally natural, we have learnt that it is not the whole story. Short-lived resonances mix effective quark and hadronic degrees of freedom.
This is in part a result of the universality of final state interactions. A key consequence of the basic principle of unitarity is that if the final state in some processes is $AB$, then the final state interaction of particles $A$ and $B$ is related to that for all $AB$ systems however they are produced. The relation is particularly simple in the region of elastic unitarity, where for instance the interactions of a low mass $\pi\pi$ system with a given spin and isospin produced in any reaction have the same phase as that in  $\pi\pi\to\pi\pi$ in the same partial wave. It is this {\it universality} of final state interactions that has to shape the QCD interactions in  decay processes. This inevitably relates quark and gluon degrees of freedom contemporaneously to those of hadrons.

\begin{figure}
\centering
 \includegraphics[width=8.cm]{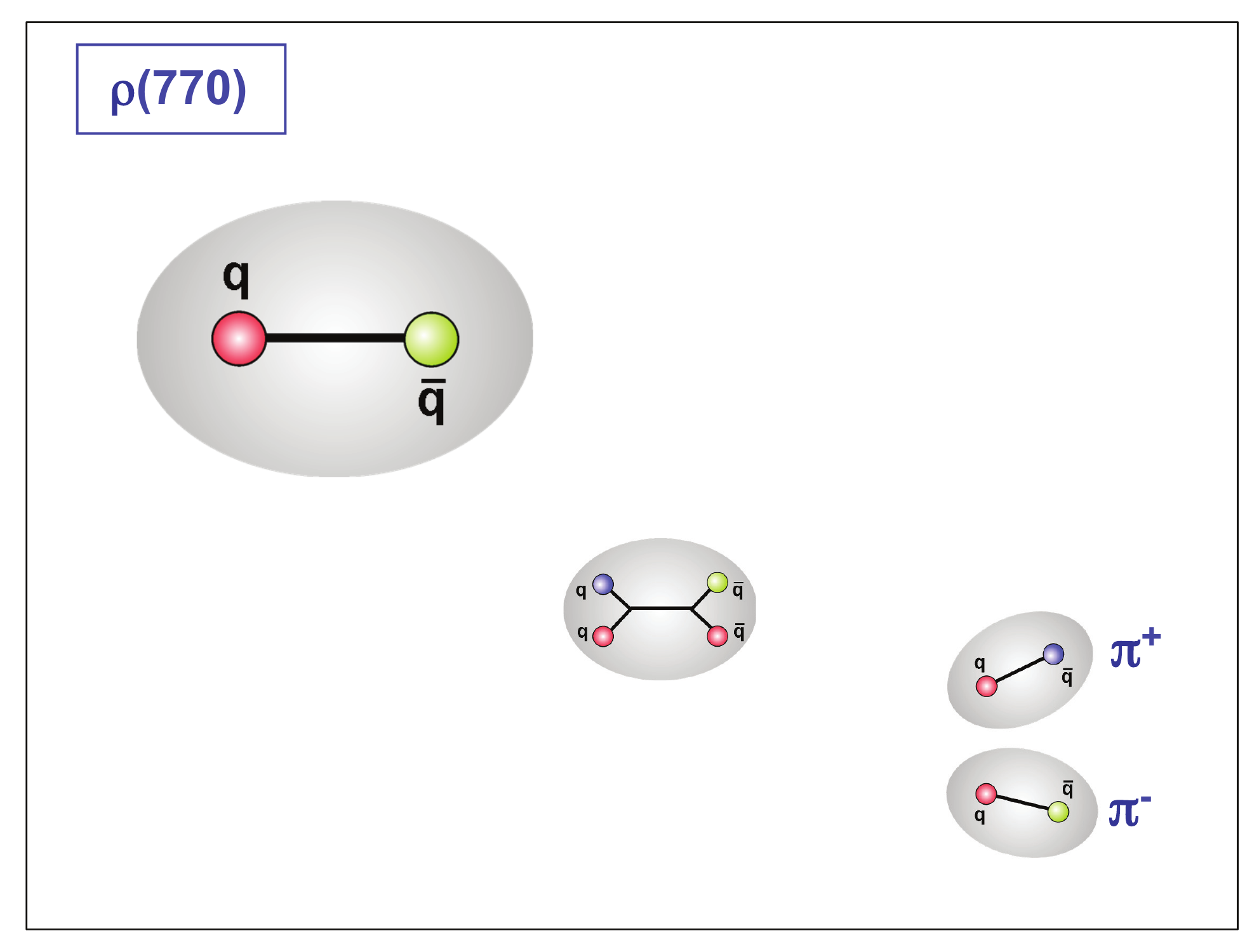}
\caption{Cartoon  of the Fock space decomposition of the $\rho$ meson at rest~\cite{pennington2}. This illustrates how on average during its lifetime this vector meson, while  having a sizeable ${\overline q}q$ component, also has smaller four quark and di-pion components, which are crucial to its eventual decay. The analogous picture for an excited baryon would naturally include pentaquark and meson-baryon components as well as a three quark combination.  }
\label{fig-13}       
\end{figure}

As soon as one has more degrees of freedom than just ${\overline q}q$ for a meson, even if the additional components are a small part of their \lq\lq wavefunction'', there may naturally arise states with orthogonal composition in which the multiquark and di-hadron components dominate. This is an idea that was realised in the calculations of van Beveren {\it et al.}~\cite{vanbeveren} and of others later~\cite{tornqvist,boglione-mrp}, that have illustrated how the nine lightest scalar mesons, which are particularly short-lived, may be dominated by di-meson components, even if their are {\it seeded} by the presence of heavier ${\overline q}q$ components~\cite{pennington2}. While experiment is fixed in a world of 3 colours, theory has no such constraint. By using Chiral Effective Field Theory or computing in lattice QCD, one can consider the world with different numbers of colours and learn how the Fock space decomposition changes (like that in Fig.~13). The ${\overline q}q$ components become narrower, while the multiquark components merge into a hadron continuum. 
The light scalars decay away as $N_c$ grows from 3, though each likely has a ${\overline q}q$ seed that is the only part remaining as $N_c$ grows large ({\it e.g.} $> 6$)~\cite{pelaez,pennington-pelaez}. Dynamical generation in baryon channels has also been studied extensively, for instance ~\cite{lutz}. These all point to hadronic components in addition to quarks as the degrees of freedom in resonances.

In the baryon sector, it had long been known that at least two states do not really fit with simple quark model expectations. The $N^*(1440)$, known as the Roper, an excitation of the nucleon, has a much lower mass than model expectations. It has strong decays to $N\pi$ and $N\pi\pi$. Indeed its Fock space image, the baryon analogue of  Fig.~13, naturally includes  pentaquark configurations that populate the $N\pi$ decay modes~\cite{riska}. So if within an effective theory of hadron interactions one thinks of the quark model as giving the \lq\lq bare'' (long-lived) states and one then corrects this with hadron interactions, the Fock space becomes more complicated. The mass of the excited state shifts from its {\it bare} position. Those with stronger couplings shifting most. Indeed, in the Effective Field Theory of \cite{julia-diaz,sato-lee-roper}, the bare state that becomes the Roper is up at 1900 MeV. How this is related to the workings of QCD is at present quite unclear.
Equally puzzling is the long known $\Lambda(1405)$ from 40 years ago. Dalitz~\cite{dalitz1} studied this over many years to see if this was a ${\overline K}N$ bound state. In the last few years 
   analysis has shown that experiment requires two states very near to each other~\cite{oset,meissner-mai}: one with the appearance of a pre-dominantly three quark baryon and the other a ${\overline K}N$ bound state.

These puzzling states always seemed \lq\lq outliers'' --- beyond the norm. However, over the past decade this perception is
 changing. As soon as one understands that the Fock space of a hadron
is not just ${\overline q}q$ or $qqq$, but there are more degrees of freedom, then it becomes clear that there may well be many hadrons (in some sense orthogonal to the picture in Fig.~13) in which the components with minimal quark number may be small. The light scalars and one of the $\Lambda(1405)$s may not be so unusual. This idea has been given a major boost, by remarkable discoveries of a whole new spectroscopy.  
As already remarked, heavy quarkonium systems of charm and bottom, for long appeared particularly simple,  given by the potential model shown in Fig.~7, but nowadays replaced by more sophisticated Effective Field Theory treatment to very good accuracy~\cite{brambilla}.

The first of the new states was the $D_{s0}^*(2317)$~\cite{palano} that is supernumerary in the charmed meson spectrum. Narrow and sitting close to the $DK$ threshold has long been thought to be a tetraquark state or a molecule. The $D_{s1}^*(2460)$ may be of similar type.
More surprises came in the hidden flavour sector as the spectrum above the open charm and beauty thresholds was studied, particularly by the $e^+e^-$ colliders at SLAC with the BaBar detector, at KEK with Belle and at BEPC with the BES detectors. As the energy increased, narrow $1^{--}$ $\psi$ states were expected to give way to higher and broader $\psi(nS)$ states.  The opening of the  ${\overline D}D$,  ${\overline D}D^*, \cdots\,$ channels shortened their lives and even shifted their masses from the predictions of a simple potential model. The surprises started in studies of $B\to KX$, where $X\to J/\psi\pi\pi$. This revealed the $X(3872)$,  some 140~MeV above the ${\overline D}D$ threshold but living 50-100 times longer than expected~\cite{belle-x3872}. Indeed, this state sits almost exactly at $D^{*0} {\overline D^0}$ threshold. Being so narrow and lying 9.5~MeV below the corresponding charged channel, isospin is violated. This is  confirmed by the appearance of the $X(3872)$ in both $J/\psi \rho$ and $J/\psi \omega$ decay modes. Analysis performed by LHCb shows that its $J^{PC}$ quantum numbers are $1^{++}$~\cite{lhcb-x3872}.

\begin{figure}
\centering
\includegraphics[width=12.cm]{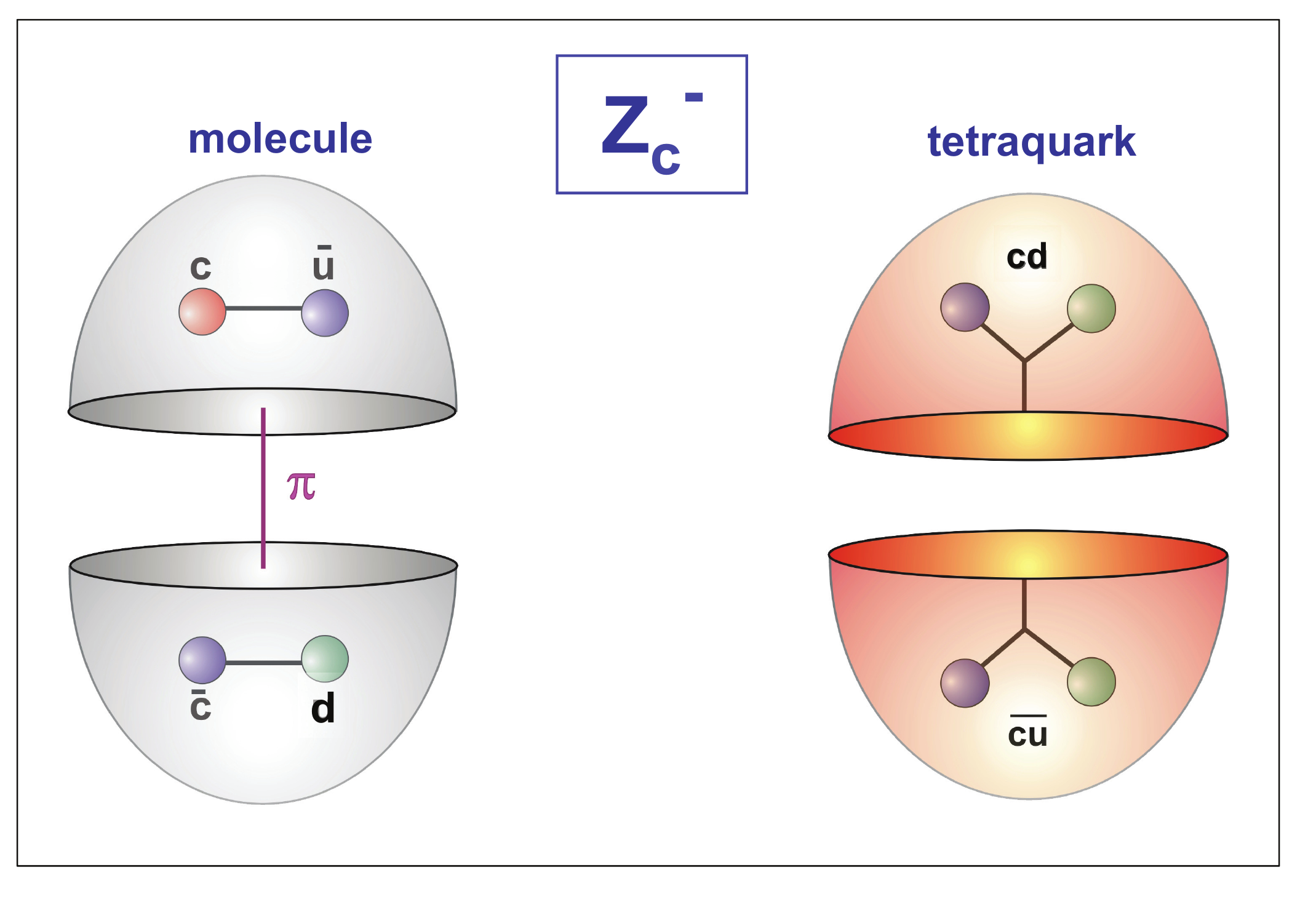}
\caption{$Z_c(4260)^-$: illustration of two possible arrangements of the minimal number of quarks ${\overline c}c{\overline u}d$ implied by its observed decay to $J/\psi\pi^-$. These are either in the form of a molecule of two colour singlet charmed mesons, bound by meson exchange, or a tetraquark state of  a diquark and an anti-diquark.     }
\label{fig-14}       
\vspace{-0.2cm}
\end{figure}

This is not the only unexpected state. Data from Belle, BaBar and BES have revealed more than a dozen $X,Y,Z$ states in both the hidden charm and bottom sectors~\cite{mitchell}. Almost all have close connections to nearby hadronic thresholds to which they couple strongly in an $S$-wave. Exactly the place where dynamically generated molecular or multiquark objects would be most favourably be generated. Indeed, the light scalars, the $a_0$ and $f_0(980)$ close to ${\overline K}K$ threshold may be exactly of this type. That more than a minimum number of quarks are required is provided by the discovery of charged charmonium-like states. The $Z_c(4260)^\pm$ must have at least four-quarks. How, these are arranged, Fig.~14, is a matter of much speculation: molecule, tetraquark or hadro-quarkonium ( a ${\overline c}c$ core with a bound meson cloud). 
Added to this, in the past year LHCb has with enormous statistics and precision studied $\Lambda_b$ decays, in which $\Lambda_b\to K^-(J/\psi p)$. In the $J/\psi p$ spectrum two new states have been identified, one broad, the other narrower, with opposite parity at 4380 and 4450~MeV respectively~\cite{aaij}. These  have been claimed to be pentaquark states, with minimal ${\overline c}uudc$ composition.

More experiments are required to expose the true nature of all these states and how they may be related. Some of the new mesons are connected by pion or radiative transitions~\cite{mitchell}. Whether they have higher spin partners or not is as yet unknown. Further data from LHCb and BESIII, and future facilities such as  Belle II and PANDA will all be required~\cite{lutz-emmi}. We have entered a new territory of QCD. Rather like the first sightings of land by Columbus and his crew, we don't yet know whether we have found a new connection to land already known, or a whole continent of hadron physics is yet to be revealed. 
That is for the next 10-20 years. 

While novelty in the charmonium sector was unexpected, that among the lighter flavours had long been suspected. The quark model neatly catalogues the quantum numbers of the states matching Fig.~2. The addition of a gluon to a ${\overline q}q$ system can produce \lq\lq exotic'' quantum numbers. The lightest of these was long anticipated to have $J^{PC}=1^{-+}$. In $\pi N$ collisions producing an $\pi\eta(')N$ final state GAMS at Serpukhov and at CERN~\cite{gams} claimed to find 
a significant $1^{-+}$ partial wave in $\pi\eta$ and $\pi\eta'$. However, the broad structure around 1.4 GeV was never sure to be a $\pi_1$ resonance. VES~\cite{ves} confirmed an enhancement in this wave, but of different shape and whether a resonance or not was questionable.  A repeat at BNL of the GAMS experiment with a more elaborate detector first claimed a resonant $\pi_1$ state at 1440 MeV and another at 1650 MeV~\cite{bnl-e852}. Subsequent analysis showed the first was an artefact of leakage from higher waves into lower~\cite{dzierba}.  Production of $3\pi$'s by COMPASS at CERN with tens of millions of events has allowed analyses of far greater sophistication~\cite{compass-spectrum}. They have shown  how the robust determination of small partial waves requires an understanding of the effect of truncating the number of contributing partial waves~\cite{haas-thesis}.  All the constraints of $S$-matrix theory from fifty years ago are required to ensure small signals can be extracted reliably and do not produce amplitudes with random moduli and phases. To learn whether new meson states are hybrids with glue as an essential constituent, or multi-quark or molecular requires the identification of not just one state but a whole flavour family. States with strangeness are crucial in confirming their composition as the flavour partners help to signal their relevant internal degrees of freedom. These are the challenges for the forthcoming high statistics and high precision experiments. COMPASS is showing the way with methods and techniques relevant to BELLE, BES, LHCb,
and the JLab experiments, GlueX and CLAS12, as well as J-PARC and PANDA@FAIR to come. Their results must wait for the next decadal anniversaries. 

\newpage
\section{Lattice QCD: calculating the spectrum and scattering}

A calculable version of the theory of the strong interaction is provided by Lattice QCD (LQCD for short). A four dimensional lattice construct makes the calculation of correlations tractable.  Studies of the excitation spectrum of quark-antiquark and three quark operators with different quantum numbers have come a long long way since first proposed by Ken Wilson forty years ago~\cite{ken-wilson}. Computing capacity has dramatically increased, as has the development of increasingly sophisticated and efficient algorithms. So despite the lattice being a discretisation of the space-time continuum, and not having the full rotational symmetry, the spectrum of excited hadrons with different spins and parities can with suitable algorithms be extracted. To make the computations in a finite volume possible, the lightest quarks are not so light. In fact often 10--20 times heavier than reality. Nevertheless an excited spectrum is deduced, see for instance ~\cite{dudek}, Fig.~15, and this accords rather closely with the structure and patterns of the simple quark model, but with the masses shifted up by the pion being at say 400 MeV, rather than its physical value of 140 MeV. Indeed it looks rather like the spectrum we see in Fig.~2.  Analogous calculations for $qqq$ baryons obtained earlier~\cite{edwards} also accord with the $SU(6) \otimes O(3)$ patterns of the $qqq$ quark model. However, there the confirmed experimental spectrum is more sparse (see discussion in Sect.~3.1).

\begin{figure}[bh]
\centering
 \includegraphics[width=14.5cm]{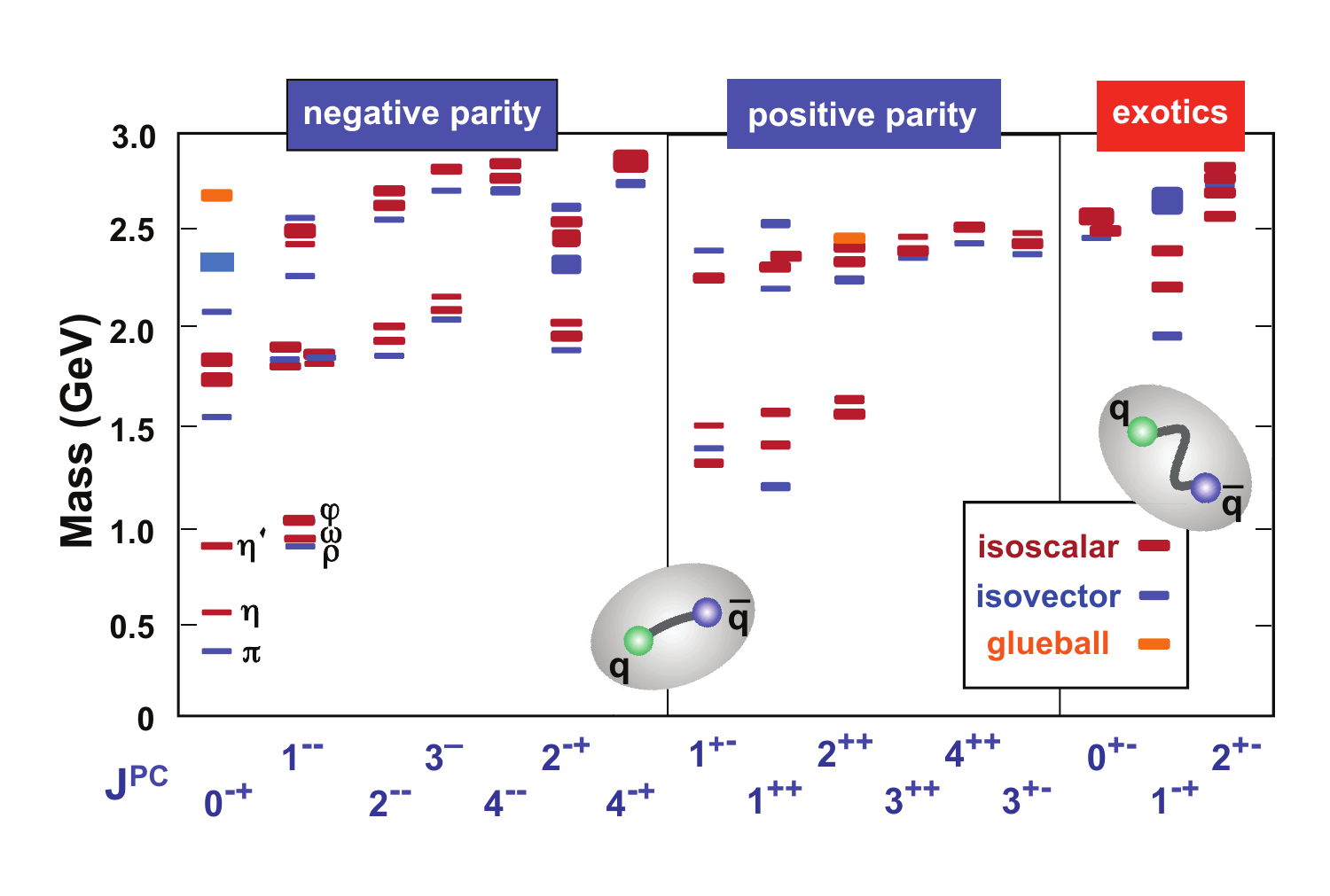}
\caption{Meson spectrum calculated in Lattice QCD by the Hadron Spectrum Collaboration~\cite{dudek}  using a wide range of operators categorised into those with conventional ${\overline q}q$ $\,J^{PC}$  quantum numbers and those that are exotic in the quark model. The calculations are with a pion mass of $\sim 400$ MeV, and do not  yet reflect the full effect of decays. Nevertheless, they point to the lightest $1^{-+}$ state being in the range of energies to be explored in detail by GlueX at JLab, and accessed by COMPASS at CERN.}
\label{fig-15}       
\vspace{-0.2cm}
\end{figure}

The inclusion of gluon operators adds more states: states with conventional quantum numbers about a GeV heavier than the corresponding  ${\overline q}q$~\cite{dudek} (Fig.~15) or  $qqq$ states~\cite{hybrid-baryons}. In the meson sector, these gluon fields can generate states with new quantum numbers, not accessible in the quark model, shown on the right in Fig.~15. The lightest are in the mass range explored already by GAMS~\cite{gams}, VES~\cite{ves}, BNL-E852~\cite{bnl-e852},  and most recently by COMPASS~\cite{compass-spectrum} in hadroproduction, and are the focus
 of GlueX and CLAS12 in photoproduction with both real and virtual photons, as just discussed in Sect.~3.  However, in calculations with single hadron operators  these states are almost stable.
Dynamical quarks do allow the creation of additional $q{\overline q}$ pairs. However these appear to have a small effect, in part because decays with 400~MeV pions  are inhibited. Worse still is in the baryon sector where the first recurrence of the nucleon with $J^P = 1/2^+$ is always hundreds of MeV too heavy~\cite{edwards}: perhaps remedied by explicit $N\pi$ and $N\pi\pi$ operators?

In the last few years remarkable progress in implementing L\"uscher's formalism~\cite{luescher} relating calculations in a box at Euclidean momenta to scattering amplitudes in Minkowski space has dramatically changed the landscape of what can be computed on the lattice. Scattering amplitudes for $\pi\pi$ and $\pi K$ have so far been extracted with coupled channel final states in each case. Such calculations demand 
the inclusion of explicit (colour neutral) di-hadron operators. As anticipated in our discussion of decays in Sect.~3.1, these are much more efficient in turning excited states into resonances: hadrons that look like those observed in experiment. Then the $\rho$ can be seen not  just as an excited state, but as a resonance with realistic couplings to $\pi\pi$. Thus the $I=1$ $P$-wave phase-shift has been determined in coupled $\pi\pi$ and $K{\overline K}$ channels, with results from the Hadron Spectrum Collaboration~\cite{dudek-wilson} being shown in Fig.~16 with pion masses of 391 and 236~MeV. The $S, P$ and $D$-waves in coupled $\pi K$ and $\eta K$ scattering
have similarly been computed. Many more channels will follow in the next few years. Real physics \lq\lq experiments'' are being performed on the lattice! The first such  studies of charmonium systems in LQCD show  that in the $J^{PC} = 1^{++}$ channel while a $\chi_{c1}$  is generated,  a state, like the $X(3872)$ close to $D^*{\overline D}$ threshold does not appear unless explicit $D^* {\overline D}$,
${\overline{D^*}}D$ and $J/\psi\omega$ operators are included~\cite{prelovsek}. To be robust, the effects of multi-body channels like $J/\psi \pi\pi$ must be included too.   Such effects can be benchmarked more readily in the light quark sector,  by including a far larger set of quark and gluon operators.

\begin{figure}
\centering
 \includegraphics[width=8.5cm]{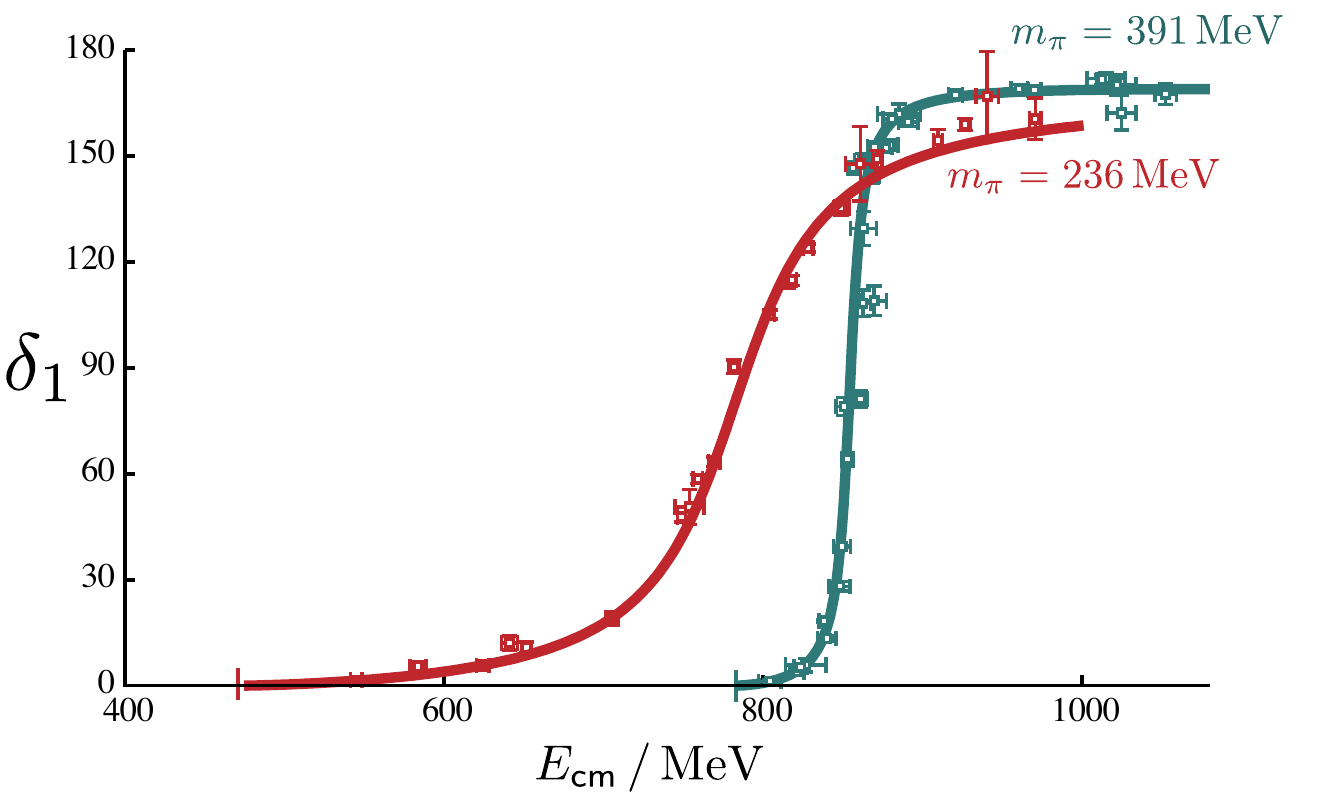}
\caption{$P$-wave $\pi\pi$ scattering phase shift, $\delta_1$ in degrees, as determined by the Hadron Spectrum Collaboration~\cite{dudek-wilson}, in worlds with different pion masses: 391 and 236 MeV. These results highlight how lattice QCD is able to compute scattering amplitudes in which a short-lived resonant state occurs, a state to be identified with the $\rho$. Fitting these scattering results with analytic functions reveals how the pole position of the $\rho$ changes with pion mass,  having a shorter lifetime  for a lighter pion. The coupling to $\pi\pi$ varies little with $m_\pi$.   }
\label{fig-16}       
\end{figure}

Such studies, {\it e.g.}~\cite{dudek-wilson} show that while the mass of the resonance and its width depend on the pion mass used, the coupling (at least for vector mesons) of the  $I=1$ states to $\pi\pi$ is essentially independent of the volume and of the light quark mass: a coupling that accords with that from experiment. Whether this is a particularity of  vector states is as yet unclear. Only by computing other systems would one know whether this equally applies to other quantum numbers. For hybrids with exotic $J^{PC}=1^{-+}$, for instance, predicting their couplings to decay channels may be key to their identification in hadro- and photo-production experiments, particularly if their signals are only a few percent of the relevant production cross-sections. This is the challenge for lattice computations in the next decade.

For fifty years the quark model (once called na\"ive and non-relativistic) has predicted a whole spectrum of $qqq$ baryons and ${\overline q}q$ mesons~\cite{capstick-isgur} in flavour multiplets.
To this static spectrum, decay patterns have been added~\cite{capstick-roberts}. However, the coupling to hadronic channels implied by decays not only gives the states a lifetime (and hence a width), but also generates feedback on the resonances' masses and flavour patterns. Lattice QCD calculations are replacing quark models. The naive non-relativistic quark model was the key to QCD, but the model is not QCD. Lattice QCD is QCD. Its calculations are a modelling much closer to the real theory. The modelling comes, not just in the finite lattice spacing, finite volumes and often unphysical quark masses, but in the truncated set of operators studied. The optimal set describing pions and kaons may be at hand. As a result resonances like the $\rho$ and $K^*$ are emerging. Excited baryons are still some time off. Nevertheless, the heavy pion mass makes the states very like the quark model, which is fine for those with high angular momentum couplings in their decays, but any channel with $S$-wave interactions is likely to be dramatically altered as the pion mass is reduced to its physical value. Lattice calculations with an extended range of operators predict states in which gluons are an essential ingredient of their structure, not just binding constituent-like quarks but adding constituent glue. The lightest of these in the baryon and meson sectors have masses 1--1.4~GeV heavier than without this glue. For mesons and baryons with these components (these essential set of operators) which have the same quantum numbers as those with ${\overline q}q$ and $qqq$ operators, these will inevitably mix through their common hadronic decay channels. When such mixings are \lq\lq orthogonalised'' it could be that predominantly $qqqg$ baryons with distinctive couplings might emerge making the identification of such states a possibility. That is yet another challenge for the future.
  
The formalism for studying electromagnetic transitions on the lattice, for instance $\pi \to\rho$ (a meson equivalent of the process in Fig.~3c), as a function of $Q^2$ has recently been set out~\cite{briceno1}. Indeed, the very first calculations, with rather heavy pions, have shown that such a formalism can indeed be implemented~\cite{briceno2}. Further study  is being spurred on as the experimental programme of photoproducing excited mesons and baryons gets under way at Jefferson Lab for the next decade. 

The $\pi\to\pi$ and $N \to N$ transitions in Fig.~3c yield the  spacelike electromagnetic formfactors of the pion and the nucleon. Lattice QCD has being tackling these for awhile including now more
computational intensive disconnected graphs that contribute for instance to the ${\overline s}s$ components of the nucleon's electromagnetic formfactor~\cite{kostas}. Though the present calculations are again for rather heavy pions, the lattice results have a precision an order of magnitude better than the difficult experimental extraction~\cite{happex-a4}. The computation of  moments of parton distributions, at least for valence quarks, {\it cf.} Fig.~9, have been performed.
However, new formulations and methodologies are being developed to compute the $x$ and $Q^2$ dependence of such distributions~\cite{ji2}. These may give access to the gluon contribution, $J_g$, in Eq.~(2) from QCD that experiment may not ever (or so easily) allow. The long term prospect is to compute the parton distributions at some $Q^2$ as in Fig.~9, and then their multi-dimensional generalisations. The era of lattice predictions (as opposed to confirmations) is at hand.

\newpage
\section{Strong coupling QCD: confinement, structure, spectrum and dynamics}

An alternative approach to calculating QCD in the strong coupling regime is provided by studies of the Schwinger-Dyson/Bound State (SD/BS) equations~\cite{collective-sdebse}. The Schwinger-Dyson equations are the field equations of the theory. They are an infinite set of nested integral equations relating each $n$-point Green's function to all others. Thus the quark and gluon propagators at all momenta satisfy integral equations, shown diagrammatically in Fig.~17, relating each of these 2-point functions to the interactions of gluons with quarks, and gluons with each other, both 3 and 4-point functions. In turn the 3-point functions are related to 2, 3 and 4-point functions, and so on {\it ad infinitum}. When combined with the inter-relations required by gauge invariance represented by the Slavnov-Taylor identities (the non-Abelian generalisations of the Ward-Green-Takahashi identities), and the constraints of multiplicative renormalisability, then one has a system of equations, the solution(s) of which define the theory completely at all momenta. Since the infinite system of integral  equations is intractable, truncations need to be made, for instance by making ans\"atze for the higher point functions. The best known of these is perturbation theory, where one works to some fixed order in the coupling $g$ or $\alpha_s$. Of course, such a perturbative truncation is not appropriate in the strong coupling regime. Then it is more appropriate to make an ansatz for the higher point functions.

 Over the past four decades progress has been made in studying the behaviour of the basic 2 and 3-point functions. That the solution of such a system of equations is unique has never been shown. If, as has often been done in the past, one focusses entirely on the infrared regime, where momenta are less than say 1~GeV, then there are known to be several solutions. However, the physical solution is the one that matches the large momentum behaviour delivered by perturbation theory as required by asymptotic freedom.
But which infrared solution does that?
\begin{figure}[b]
\centering
\includegraphics[width=12.5cm]{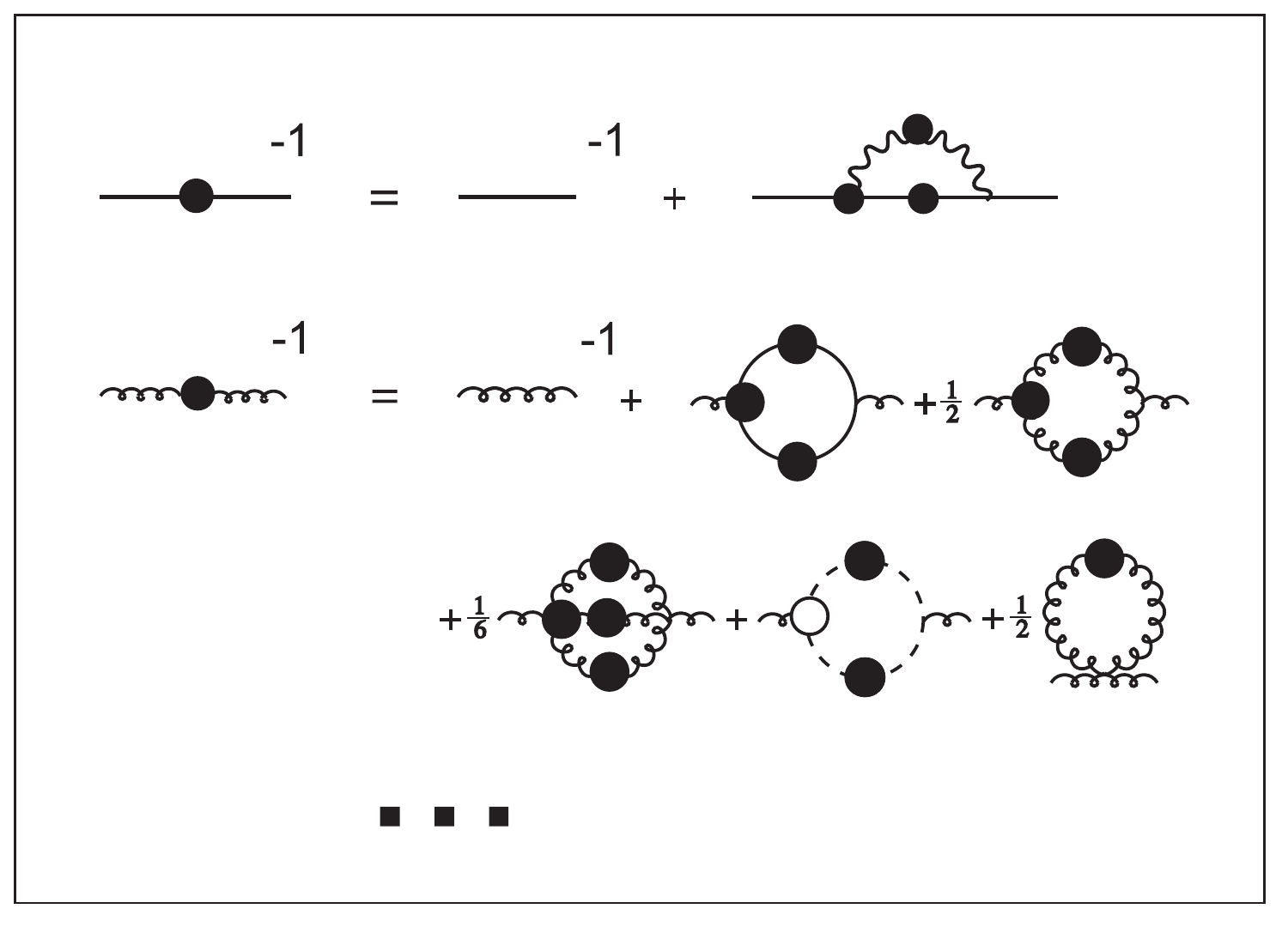}
\caption{The Schwinger-Dyson equations for the inverse quark and gluon propagators. Quarks are depicted by solid lines, gluons by curly lines, and ghosts by dashed lines. The dots on a line or vertex indicate that these are \lq\lq fully dressed'' quantities.  }
\label{fig-17}       
\end{figure}

This general approach started almost 40 years ago  studying the gluon propagator alone, by Baker, Ball and Zachariasen (BBZ)~\cite{bbz} in axial gauges, and by Pagels~\cite{pagels},  by Mandelstam~\cite{mandelstam} and by Bar-Gadda~\cite{bargadda} in covariant gauges. BBZ  found the gluon propagator became highly singular in the infrared region, with the gluon renormalisation function ${\cal G}(p^2)$, which in perturbation theory is only logarithmically different from one, behaving like $1/p^2$ for $p^2 \to 0$. Such a singular gluon would naturally give a confining force between infinitely heavy quarks with a potential linear in the interquark distance, as in Fig.~7a. However, the gluon propagator in axial gauges contains two renormalisation functions, and BBZ assumed that only the one that is non-zero at lowest order in perturbation theory need be considered, and the other ignored. Subsequently, West showed~\cite{west} that such an approximation was unjustified in gauges with positive norm states, and that such gauges could not support a singular gluon propagator. Consequently, emphasis shifted to covariant gauges where one has a simpler boson propagator, but one has to include the effect of ghosts in Fig.~17 to remove the unphysical degrees of freedom in the covariant gauge propagator. Pagels, Mandelstam and Bar-Gadda~\cite{pagels,mandelstam,bargadda}, working independently,  showed a $1/p^2$ renormalisation function could arise in the Landau gauge. Indeed Pennington and Brown~\cite{brown} performed the first numerical solution of the equations covering the whole momentum range, showing how the solution matched the expected perturbative behaviour. These studies treated the ghost and its interactions in a perturbative way.

The same gluon features in the determination of the behaviour of the quark propagator, Fig.~17. Even if the current quark is massless, the interaction is strong enough to dynamically generate a mass: a wholly strong coupling effect, Fig.~18. This is one of the truly remarkable features of this approach. Studying field theory in the continuum allows access to chiral symmetry breaking
so important for the physics of the low energy hadron world but is quite impossible to investigate theoretically either on a finite lattice or perturbatively. We return to this a little later, but first we have to deal with the treatment of ghosts. 

The unphysical degrees of freedom in the covariant gauge description of the gluon propagator are cancelled by the ghost field.
A dramatic change in the understanding of the role of ghosts followed the extensive investigation of von Smekal and the T\'ubingen group~\cite{alkofer,fischer-review}.  Again studying the infrared domain they found that the ghost and gluon behaviours were strongly coupled. In what they called the \lq\lq scaling'' solution, the gluon propagator vanished in the infrared, while the ghost became highly singular. This it was argued matched with the Zwanziger confinement conditions~\cite{zwanziger}. Watson in his thesis~\cite{watson} studied the detailed constraints on ghosts from the Slavnov-Taylor identities. Subsequently, Alkofer and Watson~\cite{alkofer-watson} showed how when the infrared behaviour dominates the Schwinger-Dyson loop integrals this could be approximated by simple power behaviour  without the need for any specific truncation scheme.   A more detailed and extensive treatment was undertaken by Fischer~\cite{fischer}. How this infrared scaling solution connected to the ultraviolet perturbative result in a way that allowed a consistent renormalisation of the transverse component of the gluon propagator required however an element of modelling of vertices and removal of divergences. 
While the behaviour of these key Green's functions in the deep infrared region is of theoretical interest, it is the momentum domain of 0.1 to 1~GeV that controls strong coupling physics.
It was never clear that simple power behaviours dominate the momentum integrals in this region. That the scaling solution does not provide the appropriate physical connection to the perturbative regime 
was prompted by a number of lattice calculations using unusually large lattices that allowed a reliable probing into the low momentum region with  studies pioneered by Cucchieri and Mendes~\cite{cucchieri-mendes} and the Adelaide group~\cite{bonnet},  with results confirmed by
M\"uller-Preussker {\it et al.}~\cite{mueller-preussker} and by Bicudo and Oliviero~\cite{bicudo-oliviero}. These studies showed a gluon propagator that became constant in the infrared, and the ghost was no more singular than $1/p^2$.

The solution of the Schwinger-Dyson equation for the gluon propagator in the Landau gauge that links to results from the lattice can be represented by an effective Euclidean mass of ~500 MeV~\cite{rodriguez-quintero,aguilar}, as had long been proposed by Cornwall~\cite{cornwall}. This generates an enhancement in the gluon propagator in the key momentum domain that drives dynamical chiral symmetry breaking. While the details of the exact structure of the dressed triple gluon vertex and of the dressed ghost-gluon interaction are still under investigation, there is now a consensus on their general behaviour and how they impact on the propagation and interactions of quarks. Phenomenological versions of these exist (if not a derivation from the QCD Lagrangian) and these are impacting on the understanding of the hadron spectrum and dynamics~\cite{roberts1}.
\begin{figure}[b]
\vspace{1mm}
 \begin{center}  
       \includegraphics*[width=12cm]{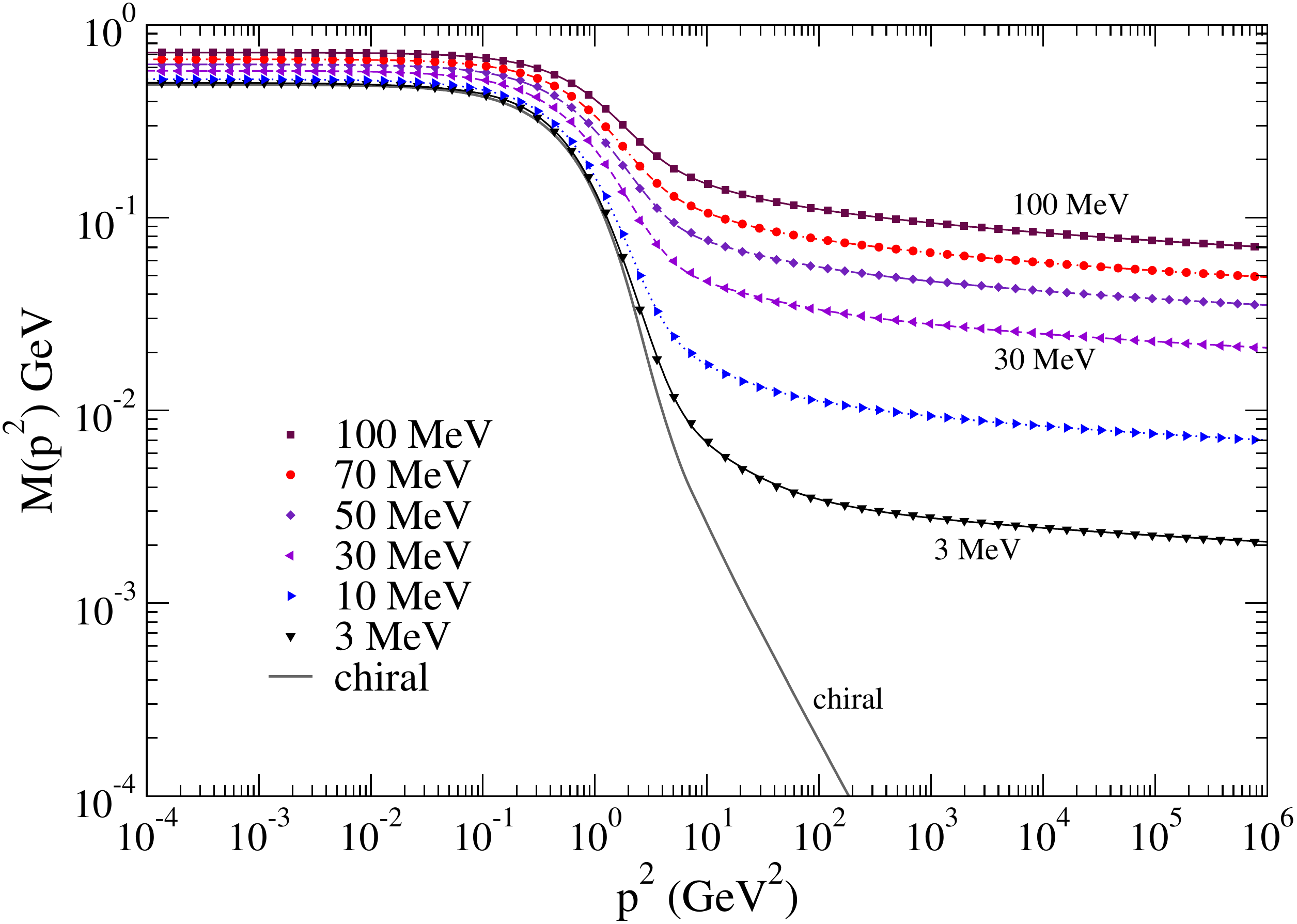}
	\caption{Euclidean mass function of a quark propagator in the Landau gauge with different current masses, specified at $p = 19$ GeV, from 0 to 100 MeV (as labelled) from Williams {\it et al.}~\cite{williams}. One sees how on a log-log plot the behaviour appears to dramatically change between a current mass of a few MeV and the massless chiral limit. This change is barely visible on a linear plot.
These results are essentially the same as found by Maris and Roberts~\cite{maris-roberts}.}
\label{fig-18}
\end{center}  
  \end{figure} 

In all these discussions of the solution for  the gluon and ghost the effect on the propagator of the light quarks has not qualitatively changed, provided  the combined effect of the gluon propagator and the quark-gluon vertex make the \lq\lq effective'' coupling strong enough in the 0.1--1 GeV momentum region to generate a significant quark mass in the infrared. The quantitative details depend on the exact gluon 2 and 3-point functions, but the general result is the same, Fig.~18.  This behaviour provides a natural interpolation between the near massless current $u$ and $d$ quarks at large Euclidean momenta to a constituent mass of $\sim 300$~MeV at momenta that are a fraction of a GeV. In the chiral limit, this behaviour seen in Fig.~18 can be fitted by the operator product expansion to show that this constituent-like mass is generated by gluon interactions that create an in-hadron ${\overline q}q$ condensate of $\sim -(250\,{\rm MeV})^3$~\cite{maris-roberts,williams}, just as deduced from experiments on $\pi\pi$ scattering from $K\to e\nu_e(\pi\pi)$ decays~\cite{na62}.
One sees that this same transition from a current quark to a constituent-like quark occurs for a range of masses, not just for $u, d$ but also the $s$ quark, adding 300--400~MeV from large to small momenta.
\begin{figure}
\centering
\includegraphics[width=13.5cm]{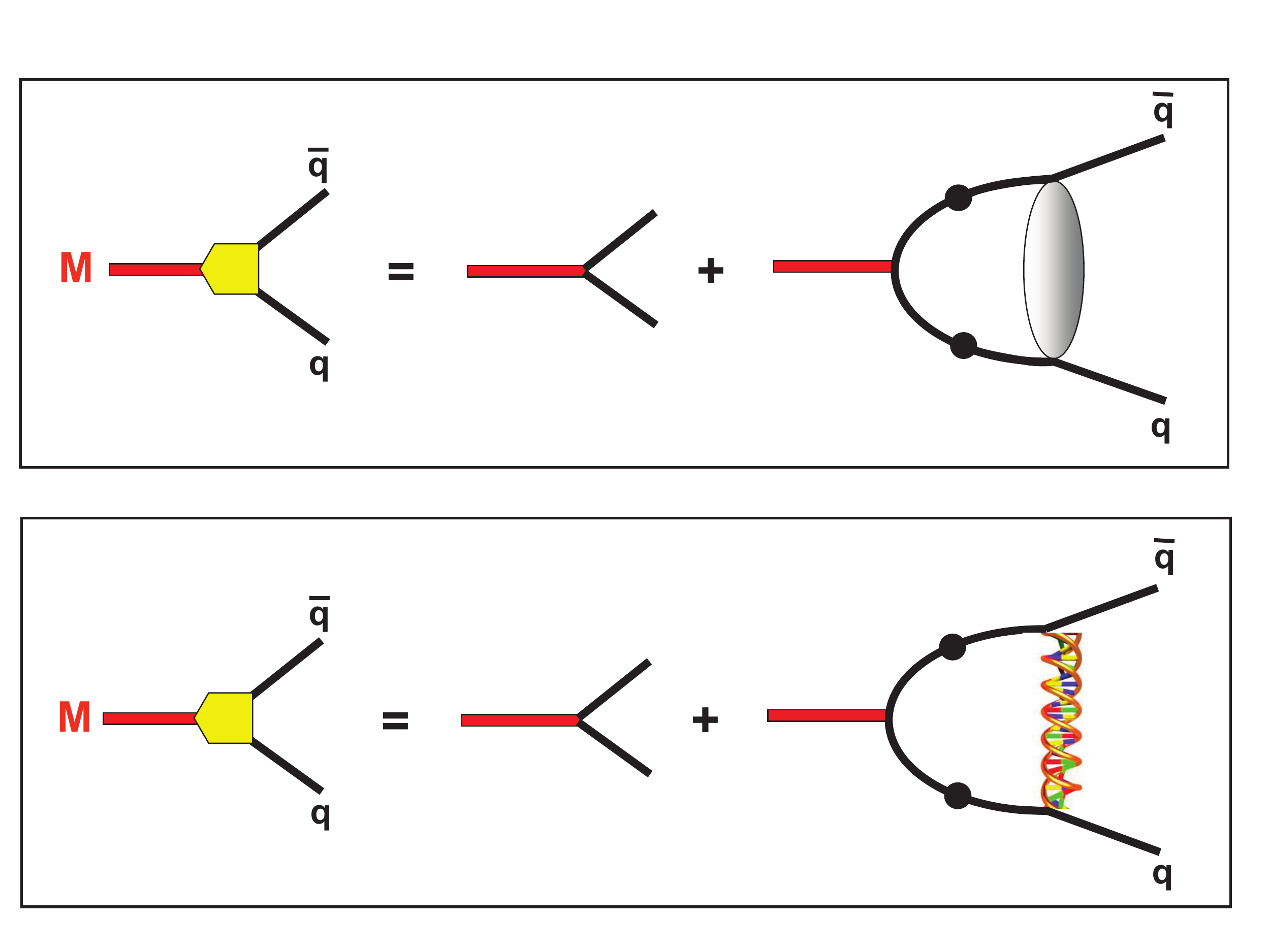}
\caption{Diagrammatic representation for the Bethe-Salpeter equation for the amplitude for a ${\overline q}q$ system. The quark propagators on the right of the equation are the same as those in Fig.~17, with results as in Fig.~18..  These quarks interact through the quark-quark scattering kernel on the right of the equation. In the Landau gauge the most used ansatz  is the rainbow ladder approximation~\cite{maristandy,cloet} illustrated in the lower figure by dressed gluon exchange. }
\label{fig-19}       
\end{figure}

In addition to explicitly showing the way quarks change from near naked current quarks in the region of asymptotic freedom to fully dressed constituent quarks when the coupling becomes strong,
the quark propagator provides the connection to physical observables through Bound State equations. In the Bethe-Salpeter equation  for mesons (Fig.~19) and  Fadeev equation for baryons, the same {\it dressed} quark propagators enter. When combined with the relevant Ward identities for the axial-vector coupling, the dynamical generation of an infrared mass for the near massless {\it up} and {\it down} current quarks, ensures chiral symmetry is broken at the hadron level: the pion is light, but its scalar partner is not.

When implemented in the Landau gauge, the rainbow ladder approximation for the quark-quark scattering kernel, illustrated in Fig.~19, appears to provide a good approximation for the ground state hadrons. The beauty of this approach is that it can treat quarks of all masses. As is well-known, the mass of the pion is not linear in the current quark mass, but rather chiral symmetry breaking dictates it grows as the square root of its mass. While pions are massless when the current quarks are massless, the masses of all other states reflect the infrared (or constituent) mass seen in Fig.~18. For instance the vector meson mass can be plotted as a function of the square of the pion mass, as the quark mass is varied from light to heavy, Fig.~20. The results from the Schwinger-Dyson/Bound State equation approach can be compared with lattice QCD calculations at heavier pion masses. In Fig.~20, we show this dependence from the Bethe-Salpeter treatment of Maris and Tandy~\cite{maristandy} and see how this provides a natural interpolation of the lattice results from  the physical $\rho$ to the $\phi$.   The corresponding results for baryons as computed using the Fadeev equation in a groundbreaking calculation by Eichmann {\it et al.}~\cite{eichmann,sanchis} for both isospin $1/2$ and $3/2$, the latter from the physical $\Delta$ to the $\Omega$ are shown in Fig.~20 too.
\begin{figure}
\centering
 \includegraphics[width=7.5cm]{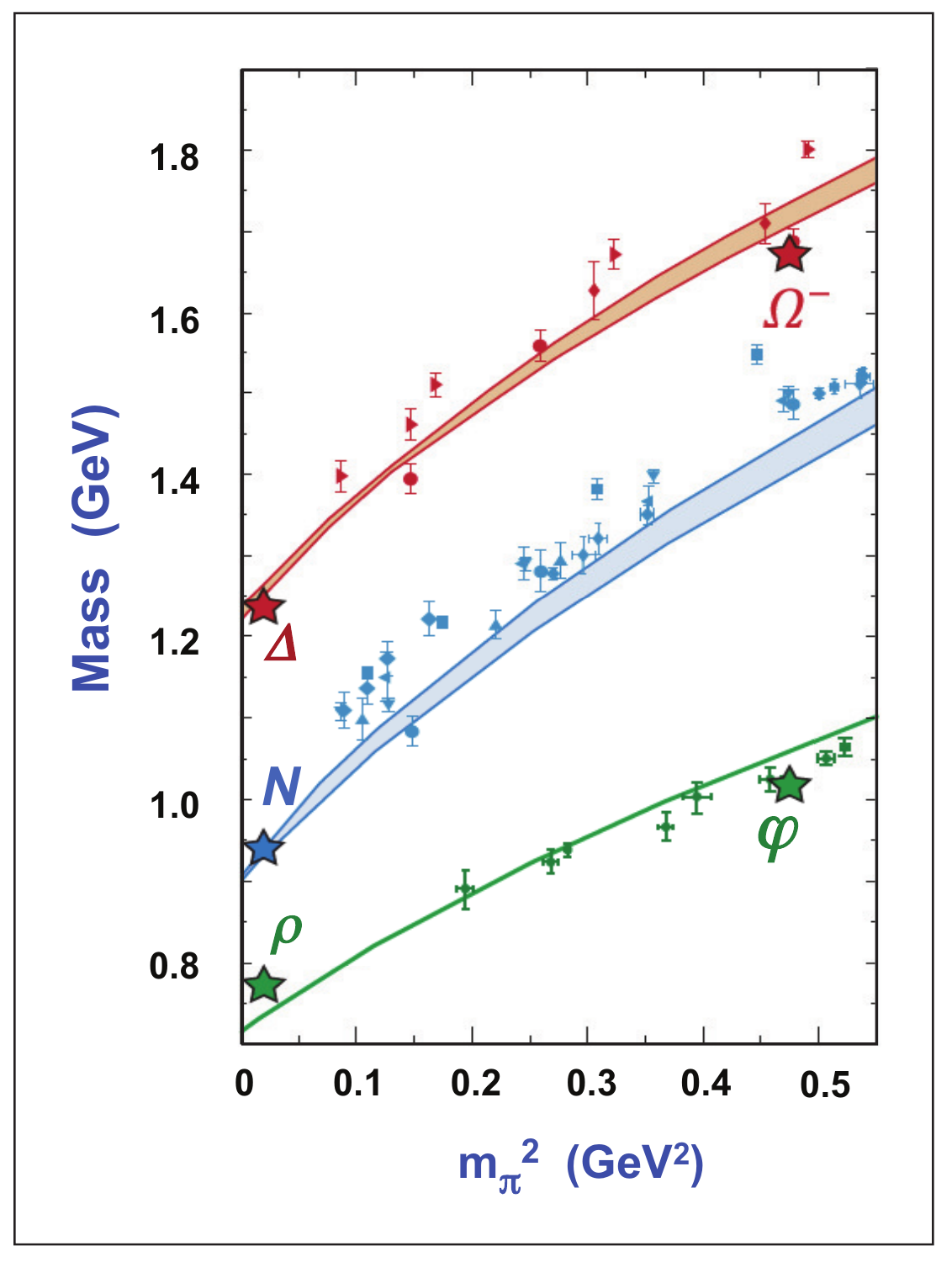}
\caption{Schwinger-Dyson/Bound State equation solutions for the masses of hadrons with different quantum numbers as a function of the square of the mass of the lightest pseudoscalar meson. Each is compared with the analogous lattice results, which are the datapoints with uncertainties. The lowest (green) curve is for vector mesons, for which two physical states are starred: the $\rho$ when $m_\pi^2 = 0.02$ GeV$^2$, the $\phi$ when $m_\pi^2 = 0.47$ GeV$^2$, {\it i.e.} an imaginary ${\overline s}s$ pseudoscalar state with mass squared = $2m_K^2-m_\pi^2$. The curve is the result from Maris and Tandy~\cite{maristandy} in the rainbow ladder approximation. The middle (blue) curve is for the $I=1/2$ nucleon state from the Fadeev equation solution found by Eichmann {\it et al.}~\cite{eichmann}. The topmost (red) curve is for the $I=3/2$ baryon from the corresponding calculation by Sanchis-Alpuis {\it et al.}~\cite{sanchis} with the physical $\Delta$ and $\Omega^-$ starred at $m_\pi^2 = 0.02, 0.47$ GeV$^2$, respectively. The lattice data are cited in ~\cite{eichmann,sanchis}.}
\label{fig-20}       
\end{figure}

While the Schwinger-Dyson approach embodies chiral symmetry breaking and can treat physical current masses for the $u$ and $d$ quarks, something that is prohibitively expensive and subject to significant finite volume corrections in lattice QCD calculations, the extension to excited states is not without difficulties. Simple approximations work less well. The rainbow ladder truncation is less accurate for quantum numbers other than vectors and pseudoscalars in the meson sector, for instance. Decays of mesons and baryons by the strong interaction that we have highlighted here are the signature of excited hadrons are not yet included in any systematic way. The lattice as we have emphasised has a path to treat this physics, even for multiparticle final states. Nevertheless a wide range of relevant results from the lattice may still be a decade away. This  equally applies to hadron structure on the lattice. Ideas on how to go beyond the first few moments of parton distributions have yet to be shown to be calculationally feasible.
  
In contrast, the SD/BS  approach can yield results more quickly. The approximations may be less under control, but the methodology is more straightforward to apply. As illustrated in Fig.~20, Bound State equation results can not only interpolate but extrapolate both lattice and experimental  data. This provides a  valuable synergy between these methodologies, helping to validate the approximations made in each. The SD/BS  approach naturally connects the strong coupling regime to that of asymptotic freedom as highlighted by the behaviour of the running quark masses shown in Fig.~18.
Thus it  is well suited to making contact with experimental results on the  electromagnetic formfactors of hadrons, for instance, and addressing the longstanding issue of when the asymptotic results from perturbation theory are relevant. The perturbative predictions for formfactors for large photon momenta of modulus  $Q$, whether spacelike or timelike, have long been known, but how  quickly are these approached.  While for inclusive hadronic quantities ``asymptotia'' seems to set in surprisingly soon just beyond  a few GeV, that is not the case for exclusive quantities. In deep inelastic scattering on a proton, when the probing photon gives one  of the nucleon's quarks a big kick, the inclusive cross-section does not seem to care how the scattered and spectator quarks arrange themselves into any particular hadronic final state. Only the sum matters. However, in the case of the nucleon's electromagnetic formfactor, the probing photon excites a proton, but it has to remain a proton. The internal dynamics takes longer to become perturbative, being governed by very many softer processes. The Schwinger-Dyson/Bound State equation approach~\cite{segovia} makes predictions about the behaviour of the pion and the nucleon formfactors out to $Q^2$ of 10-20 GeV$^2$, where experiments at Jefferson Lab will  measure these.
Similar predictions for the electromagnetic transitions from the nucleon to some excited baryons can also  be made --- an instructive quantity that has been and will continue to be studied in experiment.
Excitations with definite quantum numbers can only be extracted from data at relatively low energies, where neither the Fock space of the nucleon nor its excitation are well approximated by free quarks, but rather a complex combination of valence quarks and colour neutral modes. The latter may be pictured as the nucleon surrounded by a pion cloud at the lower $Q^2$. It is the transition from $Q^2$ of 20 GeV$^2$ down to 2--3 GeV$^2$ that reveals the complexity of the hadron, where QCD in the strong coupling regime is at work.
 Within this same SD/BS framework more differential quantities have also be studied and Generalised Parton Distributions predicted too~\cite{moutarde}: properties of which LQCD calculators can only dream at present.  

\newpage
\section{The future}
 Forty years ago with the \lq\lq confirmation'' of the quark constituents of hadrons and the \lq\lq revelation'' of partons  moving freely inside the femto-universe it seemed that hadron physics would soon be understood. Cold QCD would be a solved problem, and one could move on to search for new states of matter, like  the quark-gluon plasma. However, over the past forty years we have discovered that colour confinement works in mysterious ways, bringing a surprising richness to  hadron physics even when cold.  
At some instance one can picture a proton with  millions of gluons and perhaps a 1003 quarks and a 1000 antiquarks all sharing its momentum and spin. These partons are close to the \lq\lq bare'' (technically the renormalised) degrees of freedom in the QCD Lagrangian. Yet at the same time, the longer distance properties of the proton reflect the clustering of these many quarks and gluons under strong coupling QCD into a smaller number of {\it effective} degrees of freedom that include what we know as  constituent quarks. Unfolding experiments will confirm whether there is an analogous constituent gluon that brings new quantum number configurations to the hadron spectrum. 

A window on the link between the  multi-dimensional parton picture and the constituent image is provided by the decays and electromagnetic transitions of unstable excited hadrons. These are teaching us that the effective degrees of freedom inevitably combine coloured and colour neutral modes as the hadron evolves from creation to decay. This ensures the hadronic products emerging from the femto-universe interact in a way that is universal, with features that are independent of the way they are produced, but only depending on their overall quantum numbers.  Colour confinement means that hadron interactions both 
in experiment and in theoretical calculations  share a common language imposed by $S$-matrix principles. From lattice calculations is emerging the 21st century paradigm for the spectrum of {\it Resonances in QCD} replacing fifty year old quark models. At the same time whole galaxies of novel hadrons, $X,Y,Z$ and $P_C$'s, are being revealed by experiments --- creating new challenges and opportunities for learning about QCD. 

Experiments of unprecedented precision  and accuracy with BES III, Belle II, COMPASS, J-PARC, LHCb and all four halls at Jefferson Lab, with PANDA to come in the future, will focus on both the spectrum  and multi-dimensional structure of hadrons. For these experiments to impact on unravelling the way strong coupling QCD really works, from light to heavy flavours, demands a close synergy with theoretical approaches.  Understanding dynamics requires a combination of methodologies, from  Effective Field Theories to Schwinger-Dyson/Bound State equation treatments  and lattice QCD,  each with its own approximations and limitations. With such synergies, the planned experiments will make the next twenty years an age of discovery. The multi-dimensional structure of the proton, together with the study of the hadron spectrum, will at last reveal how from the simplicity of an underlying theory of quarks and gluons, the fascinating world of hadrons emerges by the magic of colour confinement. 

Till then may the Force be with us!

\vspace{1.cm}

\centerline{\bf Acknowledgements}
I acknowledge all the many researchers, from those just starting on the path of exploration to those with greyer hair, who have taught me much and shaped my world view over the past four decades. I also wish to thank Nobuo Sato for his construction of Figure~8, and Joanna Griffin, Pauline Russell and Shannon West  for help with many of the other figures. 
This material is based upon work supported by the U.S. Department of Energy, Office of Science, Office of Nuclear Physics under contract DE-AC05-06OR23177.


\newpage
\section{References}
\begin{thebibliography} {99}
\bibitem{yukawa} H.~Yukawa, 
Proc. Phys.-Math. Soc. Japan, {\bf I7},  48 (1937). 
(A
\bibitem{powell}
C.~M.~G.~Lattes, H.~Muirhead, G.~P.~S.~Occhialini and C.~F.~Powell.
Nature {\bf 159}, 694 (1947).
\bibitem{rochester}
G.~D.~Rochester and C.~C.~Butler,
 Nature {\bf 160} 855 (1947).

\bibitem{chew}
G.~F.~Chew, {\it S-matrix theory of strong interactions}, W.A.~Benjamin (New York) 1961,
 {\it The Analytic $S$ Matrix: a basis for nuclear democracy}, W.~A.~Benjamin (New York) 1966;
 
 G.~F.~Chew and A.~Pignotti,
  Phys.\ Rev.\  {\bf 176}, 2112 (1968);

 
  G.~F.~Chew,
  Phys.\ Rev.\ D {\bf 4}, 2330 (1971);

\bibitem{frautschi} 
  G.~F.~Chew and S.~C.~Frautschi,
  Phys.\ Rev.\ Lett.\  {\bf 7}, 394 (1961), {\it ibid.} {\bf 8}, 41 (1962).
 
 


\bibitem{gell-mann}
M.~Gell-Mann,  Phys. Lett, {\bf 8}, 214 (1964).

\bibitem{zweig}
G.~Zweig,  "An SU(3) Model for Strong Interaction Symmetry and its Breaking", CERN Report No.8182/TH.401.


\bibitem{veneziano} G.~Veneziano,
Nuovo Cim. {\bf A57}, 190-197 (1968).


\bibitem{eden} R.~J.~Eden,
Rev. Mod. Phys. {\bf 43}, 15 (1971).

\bibitem{elop} R.~J.~Eden,  P.~V.~Landshoff, D.~I.~Olive and J.~.C.~Polkinghorne,
{\it The Analytic S-Matrix}, (Cambridge University Press) 1966.

\bibitem{squires-collins}
P.~D.~B.~Collins and E.~J.~Squires, {\it Regge Poles in Particle Physics}, Springer Tracts in Modern Physics  Vol. 45 (Springer Verlag, Berlin) 1968.

\bibitem{bateman}
Bateman Manuscript Project: A. Erd{\'e}lyi, W. Magnus, F. Oberhettinger, and F.G. Tricomi, Higher Transcendental Functions (McGraw-Hill, New York, Toronto, London) Vol. 1 (1953), Vol. 2 (1953), Vol. 3 (1955)
\bibitem{string} {\it The Birth of String Theory}, ed. by A.~Capelli, E.~Castellani, F.~Colomo and P.~Di~Vecchia (Cambridge University Press) 2012.

\bibitem{OZI}
S.~Okubo, Phys. Lett. {\bf 5}, 1975 (1963);

G. Zweig, CERN Report No.8419/TH412 (1964);

J.~Iizuka, Prog. Theor. Phys. Suppl. {\bf 37}, 38 (1966).



\bibitem{pdg}
K.~A.~Olive {\it et al.} [PDG],
Chin. Phys. {\bf C38}, 090001 (2014).

\bibitem{johnson-jaffe}
A.~Chodos, R.~L.~Jaffe, K.~Johnson, C.~B.~Thorn and V.~F.~Weisskopf,
Phys. Rev. {\bf D9}, 3471 (1974).

\bibitem{thomas}
A.~W.~Thomas,
Adv. Nucl. Phys. {\bf 13}, 1 (1984).

\bibitem{weinberg}
S.~Weinberg, 
Phys. Rev. Lett. {\bf 19}, 1264 (1967).


\bibitem{salam} A.~Salam,
Proceedings of 8th Nobel Symposium Lerum, Sweden (May 19-25, 1968),
367 (1968).


\bibitem{brout-englert-higgs}
F.~Englert and R.~Brout,
Phys. Rev. Lett. {\bf 13}, 321 (1964);
 
P.~W.~Higgs,
Phys. Rev. Lett. {\bf 13}, 508 (1964),
Phys. Lett. {\bf 12}, 132 (1964).
\bibitem{fritzsch-gell-mann-leutwyler}
H.~Fritzsch, M.~Gell-Mann and H.~Leutwyler,
Phys. Lett. {\bf B47}, 365 (1973).

\bibitem{dglap}
V.~N.~Gribov and L.~N.~Lipatov,
Sov. J. Nucl. Phys. {\bf 15}, 438 (1972)
[Yad. Fiz.15,781(1972)];

Y.~L.~Dokshitzer,
Sov. Phys. JETP. {\bf 46}, 641 (1977)
[Zh. Eksp. Teor. Fiz.73,1216(1977)];

G.~Altarelli, and G.~Parisi,
Nucl. Phys. {\bf B126}, 298 (1977).


\bibitem{politzer-gross-wilczek}
D.~J.~Gross and F.~Wilczek, Phys. Rev. Lett.
{\bf 30}, 1343 (1973);

H.~D.~Politzer, Phys. Rev. Lett.  {\bf 30}, 1346 (1973).


\bibitem{ellis-gaillard-ross}
J.~R.~Ellis,  M.~K.~Gaillard  and G.~G.~Ross,
Nucl. Phys. {\bf B111}, 253 (1976)
[Erratum: Nucl. Phys. {\bf B130},516(1977)].

\bibitem{weinberg-chpt} S.~Weinberg, { Physica} {\bf 96A}, 327 (1979).

\bibitem{chpt} 
  J.~Gasser and H.~Leutwyler,
  Annals Phys.\  {\bf 158}, 142 (1984),
Nucl. Phys. {\bf B250}, 465 (1985).
\bibitem{chiral-practors} see, for instance, J.~F.~Donoghue, E.~Golowich and B.~R.~Holstein,
  Camb.\ Monogr.\ Part.\ Phys.\ Nucl.\ Phys.\ Cosmol.\  {\bf 2}, 1 (1992).
\bibitem{richardson}
J.~L.~Richardson,
Phys. Lett. {\bf B82}, 272 (1979).

\bibitem{jinshan} S.~Jin,
Hadron '95. Proceedings, 6th International Conference on
                        Hadron Spectroscopy, Manchester, UK, (July 10-14, 1995) pp.~113-119;

J.~Z.~Bai {\it et al.} [BES collaboration], Phys. Rev. Lett. {\bf 76}, 3502 (1996);

T.~Huang, S.~Jin, D.-H.~Zhang and K.-T.~Chao,
Phys. Lett. {\bf B380}, 189 (1996).


\bibitem{eichten}
 E.~Eichten, K.~Gottfried, T.~Kinoshita, J.~B.~Kogut, K.~D.~Lane and T.~M.~Yan,
  Phys.\ Rev.\ Lett.\  {\bf 34}, 369 (1975)
  [Phys.\ Rev.\ Lett.\  {\bf 36}, 1276 (1976)]:

E.~Eichten, K.~Gottfried, T.~Kinoshita, K.~D.~Lane and T.~M.~Yan,
  Phys.\ Rev.\ D {\bf 17}, 3090 (1978) {\it ibid} D {\bf 21}, 313 (1980)];


  E.~J.~Eichten and C.~Quigg,
  Phys.\ Rev.\ D {\bf 49}, 5845 (1994)
  [hep-ph/9402210].

\bibitem{CKM}N.~Cabibbo,
 Phys.\ Rev.\ Lett. {\bf 10},  531 (1963);

M.~Kobayashi and T. Maskawa,
 Prog.\ Theor.\ Phys. {\bf 49}, 652 (1973).
\bibitem{isgur-karl}
  N.~Isgur and G.~Karl,
  Phys.\ Rev.\ D {\bf 19}, 2653 (1979)
  [Phys.\ Rev.\ D {\bf 23}, 817 (1981)],
  Phys.\ Rev.\ D {\bf 20}, 1191 (1979).
\bibitem{capstick-isgur}
  S.~Capstick and N.~Isgur,
  Phys.\ Rev.\ D {\bf 34}, 2809 (1986)
  [AIP Conf.\ Proc.\  {\bf 132}, 267 (1985)].


\bibitem{accardi} J.~F.~Owens, A.~Accardi and W.~Melnitchouk, Phys. Rev. {\bf D87}, 094012 (2013).

\bibitem{nobuo}  Nobuo~Sato, W.~Melnitchouk, S.~E.~Kuhn, J.~J.~Ethier and A.~Accardi,
  JLAB-THY-16-2200, arXiv:1601.07782 [hep-ph]

\bibitem{cteq}J.~Botts {\it et al.} [CTEQ Collaboration],
  Phys.\ Lett.\ B {\bf 304}, 159 (1993)
  [hep-ph/9303255].
\bibitem{mrst}
A.~D.~Martin, R.~G.~Roberts. W.~J.~Stirling and R.~S.~Thorne,
  Phys.\ Lett.\ B {\bf 531}, 216 (2002)
  [hep-ph/0201127];

 L.~A.~Harland-Lang, A.~D.~Martin, P.~Motylinski and R.~S.~Thorne,
  Eur.\ Phys.\ J.\ C {\bf 75}, 204 (2015)
  [arXiv:1412.3989 [hep-ph]].
\bibitem{amanda}  H.~Abramowicz {\it et al.} [H1 and ZEUS Collaborations],
  Eur.\ Phys.\ J.\ C {\bf 75}, no. 12, 580 (2015)
  [arXiv:1506.06042 [hep-ex]].
\bibitem{color-glass}
 L.~D.~McLerran,
  Lect.\ Notes Phys.\  {\bf 583}, 291 (2002)
  [hep-ph/0104285];
 
  E.~Iancu, A.~Leonidov and L.~McLerran,
  hep-ph/0202270.
\bibitem{emc} J.~Aubert {\it et al.} [European Muon Collaboration], Phys. Lett. {\bf B123}, 275 (1983).
\bibitem{nmc} U.~Landgraf [NMC Collaboration],
  Nucl.\ Phys.\ A {\bf 527}, 123C (1991);
 
  E.~Rondio [NMC Collaboration],
  Nucl.\ Phys.\ A {\bf 553}, 615C (1993).
\bibitem{arrington} 
J.~Arrington, A.~Daniel, D.~Day, N.~Fomin, D.~Gaskell and P.~Solvignon,
  Phys.\ Rev.\ C {\bf 86}, 065204 (2012)
  [arXiv:1206.6343 [nucl-ex]].
\bibitem{ji-spin}   X.~Ji, X.~Xiong and F.~Yuan,
  Phys.\ Rev.\ Lett.\  {\bf 109}, 152005 (2012)
  doi:10.1103/PhysRevLett.109.152005
  [arXiv:1202.2843 [hep-ph]].  
\bibitem{lorce-leader}  E.~Leader and C.~Lorcé,
  Phys.\ Rept.\  {\bf 541}, 163 (2014)
  doi:10.1016/j.physrep.2014.02.010
  [arXiv:1309.4235 [hep-ph]].

\bibitem{bacchetta} 
  M.~Radici, F.~Conti, A.~Bacchetta and A.~Bianconi,
  arXiv:0708.0232 [hep-ph].
\bibitem{boglione} 
  M.~Boglione and A.~Prokudin,
  arXiv:1511.06924 [hep-ph].

\bibitem{hermes} {\it e.g.} A.~Airapetian {\it et al.} [HERMES Collaboration], 
  A.~Airapetian {\it et al.} [HERMES Collaboration],
  Phys.\ Rev.\ Lett.\  {\bf 84}, 4047 (2000).

\bibitem{prokudin-musch}
  B.~U.~Musch {\it et al.} [LHPC Collaboration],
  PoS LATTICE {\bf 2008}, 166 (2008);

  D.~Boer {\it et al.},
  arXiv:1108.1713 [nucl-th].

\bibitem{compass-structure}  C. Adolph {\it et al.} [COMPASS Collaboration]
Eur.\ Phys.\ J., {\bf C73} 2531 (2013).
\bibitem{mueller} D.~Mueller {\it et al.}, Fortsch. Phys. {bf 42}, 101 (1994).
\bibitem{ji-gpd} X-D. Ji, Phys. Rev. {\bf D55}, 7114 (1997).
\bibitem{radyushkin} A.~V.~Radyushkin, Phys. Lett. {\bf B380}, 417 (1996).
\bibitem{eic-whitepaper}
  M.~Anselmino {\it et al.},
  Eur.\ Phys.\ J.\ A {\bf 47}, 35 (2011)
  [arXiv:1101.4199 [hep-ex]];

  A.~Accardi {\it et al.},
  arXiv:1212.1701 [nucl-ex].
\bibitem{weiss-tagging} W.~Cosyn {\it et al.},
  J.\  Phys.\  Conf.\  Ser.\  {\bf 543}, 012007 (2014)
  [arXiv:1409.5768 [hep-ph]].
\bibitem{radici} M.~Radici, contribution to {\it Hadron 2015} (to be published).
\bibitem{Theta}
For instance, T.~Nakano {\it et al.} [LEPS Collaboration], Phys. Rev. Lett. {\bf 91}, 012002 (2003);

S.~Stepanyan {\it et al.} [CLAS Collaboration], Phys. Rev. Lett. {\bf 91}, 252001 (2003);

V.~Kubarovsky {\it etal.}, Phys. Rev. Lett. {\bf 92}, 032001 \& 049902 (2004).
\bibitem{DeVita:2006aaq} R.~De~Vita {\it et al.} [CLAS Collaboration], Phys. Rev. {\bf D74}, 032001 (2006).
\bibitem{lindenbaum}
S.~J.~Lindenbaum and R.~S.~Longacre,
  Phys.\ Lett.\ B {\bf 165}, 202 (1985).

A.~Etkin {\it et al.},
  Phys.\ Lett.\ B {\bf 201}, 568 (1988).
\bibitem{colangelo}  I.~Caprini, G.~Colangelo and H.~Leutwyler,
  Phys.\ Rev.\ Lett.\  {\bf 96}, 132001 (2006)
  [hep-ph/0512364].
\bibitem{pelaez-sigma}  
  J.~R.~Pelaez,
  arXiv:1510.00653 [hep-ph].
\bibitem{minkowski-ochs}  W.~Ochs,
  J.\ Phys.\ G {\bf 40}, 043001 (2013)
  [arXiv:1301.5183 [hep-ph]].
\bibitem{beck}  R.~Beck {\it et al.} [CBELSA-TAPS Collaboration],
  J.\ Phys.\ Conf.\ Ser.\  {\bf 295}, 012023 (2011).
\bibitem{thoma}  U.~Thoma,
  AIP Conf.\ Proc.\  {\bf 768}, 197 (2005)
  [nucl-ex/0501007].
\bibitem{anl-osaka} 
  H.~Kamano, S.~X.~Nakamura, T.-S.~H.~Lee and T.~Sato,
  Phys.\ Rev.\ C {\bf 88}, no. 3, 035209 (2013)
  [arXiv:1305.4351 [nucl-th]], PoS Hadron {\bf 2013}, 112 (2013)
  [arXiv:1312.2745 [nucl-th]].
\bibitem{bonn-gatchina}  A.~V.~Anisovich, R.~Beck, E.~Klempt, V.~A.~Nikonov, A.~V.~Sarantsev and U.~Thoma,
  Eur.\ Phys.\ J.\ A {\bf 48}, 15 (2012)
  [arXiv:1112.4937 [hep-ph]].
\bibitem{mrp-baryons}  
  M.~R.~Pennington,
  AIP Conf.\ Proc.\  {\bf 1560}, 11 (2013)
  [arXiv:1402.5435 [nucl-th]];
  EPJ Web Conf.\  {\bf 73}, 01001 (2014)
  [arXiv:1402.5406 [nucl-th]].
\bibitem{burkert} 
  V.~D.~Burkert,
  Prog.\ Part.\ Nucl.\ Phys.\  {\bf 55}, 108 (2005).
\bibitem{gothe} 
  P.~L.~Cole {\it et al.} [CLAS Collaboration],
  PoS QNP {\bf 2012}, 075 (2012).
\bibitem{mokeev} 
  V.~I.~Mokeev, I.~G.~Aznauryan and V.~D.~Burkert,
  arXiv:1109.1294 [nucl-ex].

  I.~G.~Aznauryan, V.~D.~Burkert and V.~I.~Mokeev,
  AIP Conf.\ Proc.\  {\bf 1432}, 68 (2012)
  [arXiv:1108.1125 [nucl-ex]].
\bibitem{barnes}  E.~S.~Ackleh, T.~Barnes and E.~S.~Swanson,
  Phys.\ Rev.\ D {\bf 54}, 6811 (1996)
  [hep-ph/9604355].

  T.~Barnes,
  AIP Conf.\ Proc.\  {\bf 717}, 625 (2004)
  [hep-ph/0311102].
\bibitem{isgur-geiger}  P.~Geiger and N.~Isgur,
  Phys.\ Rev.\ D {\bf 47}, 5050 (1993).
\bibitem{capstick-roberts}  S.~Capstick and W.~Roberts,
  Prog.\ Part.\ Nucl.\ Phys.\  {\bf 45}, S241 (2000)
  [nucl-th/0008028].

\bibitem{pennington2}M.~R.~Pennington,
  Acta Phys.\ Polon.\ Supp.\  {\bf 8}, no. 1, 9 (2015)
  [arXiv:1411.7902 [nucl-th]].
\bibitem{cdd}  L.~Castillejo, R.~H.~Dalitz and F.~J.~Dyson,
  Phys.\ Rev.\  {\bf 101}, 453 (1956).
\bibitem{vanbeveren}E.~van Beveren, T.A.\ Rijken, K.\ Metzger, C.\ Dullemond, G.\ Rupp and J.E.\ Ribeiro, {Z. Phys.} {\bf C30}, 615 (1986).

\bibitem{tornqvist}N.~A.~Tornqvist,
{Z.\ Phys.}   {\bf C68}, 647 (1995).
\bibitem{boglione-mrp}M.~Boglione and M.~R.~Pennington,
{Phys.\ Rev.\ Lett.}\  {\bf 79}, 1998 (1997).
\bibitem{pelaez}   C.~Hanhart, J.~R.~Pelaez and G.~Rios,
  Phys.\ Rev.\ Lett.\  {\bf 100}, 152001 (2008)
  [arXiv:0801.2871 [hep-ph]].
\bibitem{pennington-pelaez}  J.~Ruiz de Elvira, J.~R.~Pelaez, M.~R.~Pennington and D.~J.~Wilson,
  Phys.\ Rev.\ D {\bf 84}, 096006 (2011)
  [arXiv:1009.6204 [hep-ph]].
\bibitem{lutz} 
  M.~F.~M.~Lutz and E.~E.~Kolomeitsev,
  Nucl.\ Phys.\ A {\bf 755}, 29 (2005)
  [hep-ph/0501224];

 E.~Oset, E.~J.~Garzon, J.~J.~Xie, P.~Gonzalez, A.~Ramos and A.~Martinez Torres,
  AIP Conf.\ Proc.\  {\bf 1388}, 295 (2011)
  [arXiv:1103.0807 [nucl-th]].
\bibitem{riska} See, for instance, 
  D.~O.~Riska,
  Chin.\ Phys.\ C {\bf 34}, 1201 (2010).

  M.~R.~Pennington,
  AIP Conf.\ Proc.\  {\bf 1432}, 176 (2012)
  [arXiv:1109.3690 [nucl-th]].
\bibitem{julia-diaz}  B.~Julia-Diaz, H.~Kamano, T.-S.~H.~Lee, A.~Matsuyama, T.~Sato and N.~Suzuki,
  Chin.\ J.\ Phys.\  {\bf 47}, 142 (2009)
  [arXiv:0902.3200 [nucl-th]].
\bibitem{sato-lee-roper}   
  N.~Suzuki, B.~Julia-Diaz, H.~Kamano, T.-S.~H.~Lee, A.~Matsuyama and T.~Sato,
  Phys.\ Rev.\ Lett.\  {\bf 104}, 042302 (2010)
  [arXiv:0909.1356 [nucl-th]].

\bibitem{dalitz1} 
 R.~H.~Dalitz and S.~F.~Tuan,
  Phys.\ Rev.\ Lett.\  {\bf 2}, 425 (1959),
  Annals Phys.\  {\bf 10}, 307 (1960);

 R.~H.~Dalitz, J.~McGinley, C.~Belyea and S.~Anthony,
  In *Heidelberg 1982, Proceedings, Hypernuclear and Kaon Physics*, 201-214 and Oxford Univ. - 82-053 (82,REC.SEP.) 14p

  R.~H.~Dalitz,
  Eur.\ Phys.\ J.\ C {\bf 3}, 676 (1998).
\bibitem{oset}  L.~Roca and E.~Oset,
  Phys.\ Rev.\ C {\bf 87}, no. 5, 055201 (2013)
  [arXiv:1301.5741 [nucl-th]];


  V.~K.~Magas, E.~Oset and A.~Ramos,
  Phys.\ Rev.\ Lett.\  {\bf 95}, 052301 (2005)
  [hep-ph/0503043].


\bibitem{meissner-mai}
  M.~Mai and U.~G.~Meißner,
  Eur.\ Phys.\ J.\ A {\bf 51}, no. 3, 30 (2015)
  [arXiv:1411.7884 [hep-ph]];

  T.~Hyodo,
  arXiv:1512.04708 [hep-ph];

 
  R.~Molina and M.~Doring,
  arXiv:1512.05831 [hep-lat].
\bibitem{brambilla}  
  N.~Brambilla, A.~Pineda, J.~Soto and A.~Vairo,
  Nucl.\ Phys.\ B {\bf 566}, 275 (2000)
  [hep-ph/9907240].

 N.~Brambilla, M.~Groher, H.~E.~Martinez and A.~Vairo,
  Phys.\ Rev.\ D {\bf 90}, no. 11, 114032 (2014)
  [arXiv:1407.7761 [hep-ph]].
\bibitem{palano}
  A.~Palano,
  Mod.\ Phys.\ Lett.\ A {\bf 19}, 1327 (2004).

 \bibitem{belle-x3872} M.\ Heck [CDF \& D0 collaborations], {\it POS BEAUTY2009}, 041 (2009); 

W.\ Qian [LHCb collaboration], {\it Nuovo\ Cim.} {\bf C035}, 215 (2012); 

J.F.\ Liu {\it et al.} [BES collaboration], {\it Phys.\ Rev.}\ {\bf D82}, 074026 (2010);  

E.\ Prencipe [BaBar collaboration], {\it Chin.\ Phys.}\ {\bf C34}, 644 (2010); 

S.L.\ Olsen [Belle collaboration], {\it Prog.\ Theor.\ Phys.\ Suppl.}\ {\bf 193}, 38 (2012).
\bibitem{lhcb-x3872}
 R.~Aaij {\it et al.} [LHCb Collaboration],
  Phys.\ Rev.\ Lett.\  {\bf 110}, 222001 (2013)
  [arXiv:1302.6269 [hep-ex]].
\bibitem{mitchell}  
  R.~E.~Mitchell,
  PoS Hadron {\bf 2013}, 038 (2013).
\bibitem{aaij} R.\ Aaij {\it et al.} [LHCb collaboration], {\it Phys.\ Rev.\ Lett} {\bf 112}, 222002 (2014)
 [arXiv:1404.1903].

\bibitem{lutz-emmi} M.~F.~M.~Lutz {\it et al.},
Nucl.\ Phys.\ {\bf A948}, 93 (2016),
  arXiv:1511.09353 [hep-ph].
\bibitem{gams}
 D.~Alde {\it et al.} [IHEP-Brussels-Los Alamos-Annecy(LAPP) Collaboration],
  Phys.\ Lett.\ B {\bf 205}, 397 (1988);

  D.~Alde {\it et al.} [IFVE-Brussels-Annecy-Los-Alamos Collaboration],
  Sov.\ J.\ Nucl.\ Phys.\  {\bf 48}, 1035 (1988)
  [Yad.\ Fiz.\  {\bf 48}, 1727 (1988)].

\bibitem{ves} 
G.~M.~Beladidze {\it et al.} [VES Collaboration],
  Phys.\ Lett.\ B {\bf 313}, 276 (1993).


A.~Zaitsev [VES Collaboration],
  Nucl.\ Phys.\ A {\bf 675}, 155C (2000).
\bibitem{bnl-e852}
 S-U.~Chung {\it et al.} [BNL-E852], Phys.\ Rev.\ D{\bf 60}, 092001 (1999) [hep-ex/9902003];

 G.~S.~Adams {\it et al.},
  Nucl.\ Phys.\ A {\bf 680}, 335 (2000).
\bibitem{dzierba}
  A.~R.~Dzierba {\it et al.},
  Phys.\ Rev.\ D {\bf 73}, 072001 (2006)
  [hep-ex/0510068].
\bibitem{compass-spectrum}
F.~Krinner [COMPASS Collaboration],
  EPJ Web Conf.\  {\bf 96}, 01021 (2015)
  [arXiv:1501.00133 [hep-ex]];

  B.~Grube [COMPASS Collaboration],
  arXiv:1512.03599 [hep-ex].
\bibitem{haas-thesis}
Florian Haas, \lq\lq Two dimensional partial wave analysis of exclusive 190 GeV $\pi^-p$ scattering into the $\pi^-\pi^-\pi^+$ final state at COMPASS'', Ph.D. thesis submitted to Technische Universit\"at M\"unchen, CERN-Thesis-2013-277.

\bibitem{ken-wilson} K.~G.~Wilson,
  Phys.\ Rev.\ D {\bf 10}, 2445 (1974).


\bibitem{dudek}  J.~J.~Dudek, R.~G.~Edwards, M.~J.~Peardon, D.~G.~Richards and C.~E.~Thomas,
  Phys.\ Rev.\ Lett.\  {\bf 103}, 262001 (2009)
  [arXiv:0909.0200 [hep-ph]];

 J.~J.~Dudek, R.~G.~Edwards, B.~Joo, M.~J.~Peardon, D.~G.~Richards and C.~E.~Thomas,
  Phys.\ Rev.\ D {\bf 83}, 111502 (2011)
  [arXiv:1102.4299 [hep-lat]].
\newpage
\bibitem{edwards}  R.~G.~Edwards, J.~J.~Dudek, D.~G.~Richards and S.~J.~Wallace,
  Phys.\ Rev.\ D {\bf 84}, 074508 (2011)
  [arXiv:1104.5152 [hep-ph]].

  R.~G.~Edwards {\it et al.} [Hadron Spectrum Collaboration],
  Phys.\ Rev.\ D {\bf 87}, 054506 (2013)
  [arXiv:1212.5236 [hep-ph]].

\bibitem{hybrid-baryons} 
  J.~J.~Dudek and R.~G.~Edwards,
  Phys.\ Rev.\ D {\bf 85}, 054016 (2012)
  [arXiv:1201.2349 [hep-ph]].
\bibitem{luescher}
  M.~L\"uscher,
  Commun.\ Math.\ Phys.\  {\bf 105}, 153 (1986).
\bibitem{dudek-wilson}
 J.~J.~Dudek {\it et al.} [Hadron Spectrum Collaboration],
  Phys.\ Rev.\ Lett.\  {\bf 113}, 182001 (2014)
  [arXiv:1406.4158 [hep-ph]];

 D.~J.~Wilson, J.~J.~Dudek, R.~G.~Edwards and C.~E.~Thomas,
  Phys.\ Rev.\ D {\bf 91}, 054008 (2015)
  [arXiv:1411.2004 [hep-ph]].
\bibitem{prelovsek}
S. Prelovsek and L. Leskovec, Phys.\ Rev.\ Lett. {\bf 111}, 192001 (2013) [arXiv:1307.5172[hep-lat]].

  S.~Prelovsek,
  arXiv:1508.07322 [hep-lat].

  M.~Padmanath, C.~B.~Lang and S.~Prelovsek,
  Phys.\ Rev.\ D {\bf 92}, 034501 (2015)
  [arXiv:1503.03257 [hep-lat]].
\bibitem{briceno1} 
  R.~A.~Briceno, M.~T.~Hansen and A.~Walker-Loud,
  Phys.\ Rev.\ D {\bf 91}, no. 3, 034501 (2015)
  [arXiv:1406.5965 [hep-lat]].
\bibitem{briceno2}   C.~J.~Shultz, J.~J.~Dudek and R.~G.~Edwards,
  Phys.\ Rev.\ D {\bf 91}, 114501 (2015)
  [arXiv:1501.07457 [hep-lat]].

  R.~A.~Briceno, J.~J.~Dudek, R.~G.~Edwards, C.~J.~Shultz, C.~E.~Thomas and D.~J.~Wilson,
Phys.\ Rev.\ Lett. {\bf 115}, 242001
  [arXiv:1507.06622 [hep-ph]].
\bibitem{kostas}
  J.~Green {\it et al.},
  Phys.\ Rev.\ D {\bf 92}, 031501 (2015)
  [arXiv:1505.01803 [hep-lat]].

\bibitem{happex-a4}  K.~A.~Aniol {\it et al.} [HAPPEX Collaboration],
  Phys.\ Rev.\ Lett.\  {\bf 96}, 022003 (2006)
  [nucl-ex/0506010];

D. Androic {\it et al.} [G0 Collaboration],
Phys.\ Rev.\ Lett.\ {\bf 104}, 012001 (2010),
arXiv:0909.5107 [nucl-ex];

 F.~E.~Maas {\it et al.} [A4 Collaboration],
  Phys.\ Rev.\ Lett.\  {\bf 93}, 022002 (2004)
  [nucl-ex/0401019];

S. Baunack {\it et al.},
Phys.\ Rev.\ Lett. {\bf 102}, 151803 (2009),
arXiv:0903.2733 [nucl-ex].


\bibitem{ji2}X. Ji, Phys. Rev. Lett.
{\bf 110}, 262002 (2013) [arXiv:1305.1539 [hep-ph]];

X. Ji, J. -H. Zhang and Y. Zhao, Phys.\ Rev.\ Lett.
{\bf 111}, 112002 (2013) [arXiv:1304.6708 [hep-ph]].
\bibitem{collective-sdebse} 
  A.~Bashir, L.~Chang, I.~C.~Cloet, B.~El-Bennich, Y.~X.~Liu, C.~D.~Roberts and P.~C.~Tandy,
  Commun.\ Theor.\ Phys.\  {\bf 58}, 79 (2012)
  [arXiv:1201.3366 [nucl-th]].


\bibitem{bbz} M.~Baker, J.~S.~Ball and F.~Zachariasen 
{Nucl.\ Phys.}  {\bf B186} 531, 560 (1981).
\bibitem{pagels}H.~Pagels, {\ Phys.\ Rev.} {\bf D15} 2991 (1977).

\bibitem{mandelstam} S.~Mandelstam, {Phys.\ Rev.}  {\bf D20} 3223 (1979).
\bibitem{bargadda}U.~Bar-Gadda, {Nucl.\ Phys.}  {\bf B163} 312 (1981).
\bibitem{west} G.~B.~West, {\ Phys.\ Lett.}  {\bf B115} 468 (1982);
{Phys.\ Rev.}  {\bf D27} 1878 (1983).
\bibitem{brown}N.~Brown and M.~R.~Pennington, {Phys.\ Lett.} {\bf B202} 257  (1988) (erratum-{\it ibid.} {\bf B205} 596 (1988)), {Phys.\ Rev.}  {\bf D38} 2266 (1988);
{Phys. Rev.} {\bf D39} 2723 (1989).

\bibitem{alkofer} 

R.~Alkofer and L.~von Smekal, { Nucl.\ Phys.} {\bf A680} 133 (2000) [hep-ph/0004141];
{\it Phys.\ Rept.}  {\bf 353}  281 (2001) (hep-ph/0007355);

C.~S.~Fischer and R.~Alkofer, {Phys.\ Rev.} {\bf D67}  094020 (2003) [hep-ph/0301094].
\bibitem{fischer-review} 
  C.~S.~Fischer,
  J.\ Phys.\ G {\bf 32}, R253 (2006)
  [hep-ph/0605173].
\bibitem{zwanziger} 
  D.~Zwanziger,
  Nucl.\ Phys.\ B {\bf 364}, 127 (1991),
Phys.\ Rev.\ {\bf D65} 094039 (2002)
[hep-th/0109224],
  Phys.\ Rev.\ D {\bf 69}, 016002 (2004)
  [hep-ph/0303028].
\bibitem{watson}P.~Watson, {\it The inclusion of ghosts in Landau gauge Schwinger-Dyson studies of infrared QCD}, Ph.D. thesis submitted to Durham University, 2000,
[hep-ph/9901454] (unpublished).
 
\bibitem{alkofer-watson} 
  P.~Watson and R.~Alkofer,
  Phys.\ Rev.\ Lett.\  {\bf 86}, 5239 (2001)
  [hep-ph/0102332].
\bibitem{fischer}  C.~S.~Fischer, Ph.D. thesis submitted to the University of T\"ubingen,
  hep-ph/0304233.


\bibitem{cucchieri-mendes}
  A.~Cucchieri and T.~Mendes,
  Phys.\ Rev.\ D {\bf 73}, 071502 (2006)
  [hep-lat/0602012];
  PoS LAT {\bf 2007}, 297 (2007)
  [arXiv:0710.0412 [hep-lat]];
  PoS LATTICE {\bf 2010}, 280 (2010)
  [arXiv:1101.4537 [hep-lat]].
\bibitem{bonnet}
F.~D.~Bonnet, P.~O.~Bowman, D.~B.~Leinweber and A.~G.~Williams,
{Phys.\ Rev.} {\bf D62} 051501 (2000) [hep-lat/0002020]. 

\bibitem{mueller-preussker}
  I.~L.~Bogolubsky, E.~M.~Ilgenfritz, M.~Muller-Preussker and A.~Sternbeck,
  Phys.\ Lett.\ B {\bf 676}, 69 (2009)
  [arXiv:0901.0736 [hep-lat]].
\bibitem{bicudo-oliviero}  P.~Bicudo and O.~Oliveira,
  PoS LATTICE {\bf 2010}, 269 (2010)
  [arXiv:1010.1975 [hep-lat]].
\bibitem{rodriguez-quintero}  P.~Boucaud, M.~E.~Gomez, J.~P.~Leroy, A.~Le Yaouanc, J.~Micheli, O.~Pene and J.~Rodriguez-Quintero,
  Phys.\ Rev.\ D {\bf 82}, 054007 (2010)
  [arXiv:1004.4135 [hep-ph]].
\bibitem{aguilar}  A.~C.~Aguilar, D.~Binosi and J.~Papavassiliou,
  Phys.\ Rev.\ D {\bf 84}, 085026 (2011)
  [arXiv:1107.3968 [hep-ph]].
\bibitem{cornwall}  J.~M.~Cornwall,
  Phys.\ Rev.\ D {\bf 26}, 1453 (1982).
\bibitem{roberts1} 
  C.~D.~Roberts,
  Prog.\ Part.\ Nucl.\ Phys.\  {\bf 61}, 50 (2008)
  [arXiv:0712.0633 [nucl-th]].

\bibitem{williams} R.~Williams, C.~S.~Fischer and M.~R.~Pennington,
  Phys.\ Lett.\ B {\bf 645}, 167 (2007)
  [hep-ph/0612061].
\bibitem{maris-roberts} P.~Maris and C.~D.~Roberts, {Phys.\ Rev.} {\bf C56}, 3369 (1997). 
\bibitem{na62}  J.~R.~Batley {\it et al.} [NA48-2 Collaboration],
  Eur.\ Phys.\ J.\ C {\bf 70}, 635 (2010).
\bibitem{maristandy} P.~Maris and P.~Tandy, 
  P.~Maris and P.~C.~Tandy,
  Phys.\ Rev.\ C {\bf 60}, 055214 (1999)
  [nucl-th/9905056]. 


\bibitem{cloet}
  G.~Eichmann, R.~Alkofer, I.~C.~Cloet, A.~Krassnigg and C.~D.~Roberts,
  Phys.\ Rev.\ C {\bf 77}, 042202 (2008)
  [arXiv:0802.1948 [nucl-th]].



\bibitem{eichmann} 
D.~Nicmorus, G.~Eichmann, A.~Krassnigg and R.~Alkofer,
  PoS CONFINEMENT {\bf 8}, 052 (2008)
  [arXiv:0812.2966 [hep-ph]].

 G.~Eichmann,
  J.\ Phys.\ Conf.\ Ser.\  {\bf 426}, 012014 (2013).


\bibitem{sanchis} 
  H.~Sanchis-Alepuz, G.~Eichmann, S.~Villalba-Chavez and R.~Alkofer,
  Phys.\ Rev.\ D {\bf 84}, 096003 (2011)
  [arXiv:1109.0199 [hep-ph]].


\bibitem{segovia} 
  J.~Segovia, I.~C.~Cloet, C.~D.~Roberts and S.~M.~Schmidt,
  Few Body Syst.\  {\bf 55}, 1185 (2014)
  [arXiv:1408.2919 [nucl-th]].

  C.~D.~Roberts,
  arXiv:1509.08952 [nucl-th].

\bibitem{moutarde} C.~Mezrag, L.~Chang, H.~Moutarde, C.~D.~Roberts, J.~Rodríguez-Quintero, F.~Sabatié and S.~M.~Schmidt,
  Phys.\ Lett.\ B {\bf 741}, 190 (2015)
  [arXiv:1411.6634 [nucl-th]].
\end{thebibliography}
\end{document}